\documentclass[]{imag-ms-template}
\usepackage{float}
\usepackage{subcaption}
\usepackage{graphicx}
\usepackage{amsthm}

\title{EEGSync: Unitary Temporal Alignment Method for Naturalistic and Resting-State EEG}

\author{Anand A. Joshi,$^{1\ast\dagger}$ Satwik Bathula,$^{1\dagger}$ Richard M. Leahy$^{1}$\\
{\small $^{1}$Ming Hsieh Dept. of Electrical and Computer Engineering, University of Southern California,}\\
{\small Los Angeles, CA 90089, USA}\\
{\small $^\ast$Correspondence:  ajoshi@usc.edu}
{\small $^\dagger$ Both authors contributed equally.}
}

\begin{document} 

\maketitle


\begin{abstract}
Electroencephalography (EEG) studies increasingly use paradigms that lack discrete event timing, including resting state, movie watching, story listening, and other naturalistic experiments. These recordings contain structured oscillatory activity, but they often lack repeated stimulus onsets that define a shared time clock across trials or participants. We introduce EEGSync, a complex-valued unitary Procrustes temporal alignment method for band-limited analytic EEG signals. EEGSync represents each subject as a matrix of complex oscillatory trajectories and solves a unitary Procrustes problem that maps a target subject into the temporal coordinate system of a reference or group template. The method is computationally tractable through an economy-SVD factorization that avoids constructing the full temporal operator. This temporal alignment strictly preserves the original pairwise relationships and power across EEG channels. The unitary transformation goes beyond a simple time shift, aligning latent temporal structure rather than estimating a physical clock offset. Consequently, nonlinear instantaneous measures such as phase-locking value, amplitude-envelope correlation, and phase-amplitude coupling are not guaranteed to be invariant. This framing makes EEGSync especially suitable for naturalistic shared-stimulus data, where the stimulus provides a continuous temporal scaffold, and potentially useful for resting-state data when interpreted as covariance-preserving latent manifold alignment. 

Evaluated under strict subject-independent cross-validation across physiological and clinical EEG settings, \mbox{EEGSync}-based features achieved an AUROC of 0.9600 and a maximum classification accuracy of 89.45\% in eyes-open versus eyes-closed state discrimination, using different model configurations. For Alzheimer's disease versus healthy control classification, the highest AUROC reached 0.9272 in eyes-closed recordings, with additional discrimination observed under eyes-open photostimulation. These findings suggest that temporal alignment captures discriminative information that can complement conventional spectral and statistical EEG features.

\end{abstract}

\section{Introduction}

Many EEG and MEG analyses are designed around event timing. In an event-related potential experiment, the same discrete event is repeated many times, and averaging is meaningful because each trial has a known onset. A growing fraction of cognitive and clinical neuroscience, however, studies brain activity during continuous or weakly structured conditions like movie watching, story listening, music, free viewing, and unconstrained rest. These recordings are not stimulus-locked in the ERP sense, yet they are not all equivalent. Naturalistic experiments usually contain a shared external drive, whereas resting-state recordings do not.

This distinction is central. During a movie or story, different subjects receive the same continuous sensory input, so it is reasonable to expect a shared but subject-specific response trajectory. During resting state, there is no common stimulus time course. A method that aligns resting-state subjects can still be useful, but its output should be interpreted as an alignment of latent oscillatory state spaces rather than a recovered common waveform.

EEG temporal alignment requires care because oscillatory signals have both amplitude and phase, and instantaneous phase is meaningful only after appropriate band-limited preprocessing. These facts motivate a complex-valued formulation based on analytic signals rather than a real-valued alignment of broadband voltages. Conventional alignment strategies present notable limitations, as Dynamic Time Warping (DTW) introduces non-uniform temporal warping that distorts phase-frequency relationships, while spatial hyperalignment or Canonical Correlation Analysis (CCA) mix electrode channels across subjects, compromising native spatial topology.

We propose \textbf{EEGSync}, a unitary Procrustes framework for aligning band-limited analytic EEG across subjects. The method keeps the flexibility of a general temporal unitary matrix, but the interpretation is deliberately modest. EEGSync aligns complex trajectories in a shared latent coordinate system and exactly preserves the pairwise relationships between EEG channels. It does not, without additional assumptions, guarantee preservation of nonlinear instantaneous phase or amplitude-envelope metrics, which are instead evaluated as empirical distortion diagnostics.

EEGSync is designed to reduce temporal mismatch between EEG recordings while leaving the electrode channels themselves unchanged. The method therefore aligns temporal structure across recordings without mixing information across channels.

This work introduces EEGSync, a novel complex unitary temporal alignment framework for aligning analytic EEG recordings within a shared latent temporal coordinate system. The method admits an exact Economy-SVD formulation with computational complexity $\mathcal{O}(V^2T+V^3)$, avoiding explicit construction of the full $T\times T$ temporal operator. EEGSync also preserves the channel power and pairwise inner products $ZZ^\dagger$ exactly, maintaining the pairwise relationships between EEG channels of each recording while allowing temporal coordinates to be reparameterized. We validate the framework under strict subject-independent evaluation, achieving an AUROC of $0.9600$ for eyes-open versus eyes-closed classification and up to $0.9272$ for Alzheimer's disease discrimination.

\section{Related Work}
\label{sec:related_work}

Cross-subject EEG analysis is challenging because recordings vary substantially across participants in amplitude, phase, spatial distribution, and temporal organization. Existing approaches address this variability through spatial decomposition, covariance and domain alignment, temporal synchronization, or learned shared representations. \mbox{EEGSync} is most closely related to these methods but differs in that it aligns complex EEG recordings exclusively along the temporal dimension while leaving the original electrode coordinates unchanged.

\paragraph{Spatial Decomposition and Shared EEG Representations.}
A large body of EEG signal processing focuses on transforming the channel representation into a more informative latent space. Independent Component Analysis (ICA) decomposes multichannel EEG into statistically independent components and has become widely used for source separation and artifact removal \citep{makeig1997ica,vigario1997ica,jung2000ica,delorme2004eeglab}. Common Spatial Patterns (CSP) similarly constructs spatial filters that maximize differences in band-limited variance between experimental conditions and remains a standard approach for motor-imagery EEG \citep{pfurtscheller1999erd,ramoser2000csp}. Canonical Correlation Analysis (CCA) constructs linear projections that maximize correlation between signal representations and has been used extensively in EEG, particularly for steady-state evoked responses \citep{lin2006cca}. Multiway CCA extends this principle to jointly extract components shared across multiple neural recordings \citep{decheveigne2019multiway}. More recent neural-network approaches, such as EEGNet and related compact architectures, also learn task-specific spatial and temporal representations directly from EEG \citep{lawhern2018eegnet,wang2020accurate}. These approaches can provide highly informative latent representations, but they generally transform, combine, or learn projections of the electrode channels. In contrast, \mbox{EEGSync} retains the original electrode coordinate system and applies its transformation only to temporal coordinates.

\paragraph{Cross-Subject Covariance and Domain Alignment.}
A second line of work addresses inter-subject variability by aligning the statistical distributions or covariance structure of EEG recordings. Riemannian methods represent EEG covariance matrices as points on the manifold of symmetric positive-definite matrices and perform classification or transfer in the resulting geometric space \citep{barachant2012riemannian}. This perspective has led to several cross-subject transfer methods. Riemannian transfer approaches align subject-specific covariance distributions before classification \citep{zanini2018riemannian}, while Riemannian Procrustes Analysis combines translation, scaling, and rotation in tangent or covariance representations to reduce domain mismatch \citep{rodrigues2019riemannian}. Euclidean Alignment similarly normalizes EEG trials using subject-specific covariance information before cross-subject learning \citep{he2020euclidean}. More broadly, transfer-learning frameworks have been developed to exploit data from previously observed subjects \citep{jayaram2016transfer}, while Manifold Embedded Knowledge Transfer further combines covariance alignment with distribution adaptation in a shared feature space \citep{zhang2020mekt}. These methods demonstrate the importance of reducing inter-subject variability in EEG, but they primarily achieve alignment by modifying covariance or feature representations. \mbox{EEGSync} instead operates directly on the complex multichannel time series and preserves each recording's channel-to-channel inner-product matrix $ZZ^\dagger$ exactly.

\paragraph{Temporal Alignment and Synchronization in EEG.}
Temporal variability can also be addressed directly along the time axis. Dynamic Time Warping and related approaches locally expand or contract temporal trajectories to account for latency differences. In EEG-based brain--computer interfaces, weighted Dynamic Time Warping has been used to improve transfer of motor-imagery recordings across subjects \citep{giles2020weighted}. Such methods explicitly modify temporal correspondence and can therefore alter local timing, phase, and frequency relationships. This consideration is particularly important for oscillatory EEG, where amplitude and phase dynamics carry physiological information \citep{cohen2014}. \mbox{EEGSync} takes a different approach by representing band-limited EEG as a complex analytic signal and applying a unitary transformation to its temporal coordinates rather than explicitly warping physical time.

Shared temporal structure has also been studied directly across participants. In naturalistic EEG experiments, inter-subject correlation and correlated-component approaches have shown that common continuous stimuli can produce reproducible temporal activity across individuals \citep{dmochowski2012isc}. Multi-subject component methods similarly seek common activity patterns across recordings \citep{decheveigne2019multiway}. These approaches motivate the search for shared structure across participants but typically quantify or extract correlated responses rather than explicitly aligning entire multichannel EEG recordings within a common temporal coordinate system.

\paragraph{Relation to Representational Alignment in Neuroimaging.}
Related ideas have been developed outside EEG. Hyperalignment uses Procrustes transformations to construct a common representational space across subjects from fMRI measurements \citep{haxby2011common}. More directly related mathematically, BrainSync aligns resting-state fMRI time series using orthogonal transformations along the temporal dimension while preserving spatial correlation structure \citep{joshi2018brainsync}. These methods provide important conceptual precedents for coordinate alignment in neural data, but they were developed for real-valued, low-frequency fMRI signals rather than oscillatory electrophysiology.

\paragraph{Positioning of EEGSync.}
\mbox{EEGSync} combines aspects of these lines of work while addressing a different alignment problem. Unlike ICA, CSP, CCA, and covariance-based transfer methods, it does not construct a new spatial basis or mix electrode channels. Unlike Dynamic Time Warping, it does not impose pointwise correspondence through local expansion or contraction of physical time. Instead, \mbox{EEGSync} represents EEG using complex analytic signals and computes a unitary transformation acting strictly along the temporal dimension. This allows temporal variability to be reduced while exactly preserving channel power and pairwise channel inner products. The resulting framework provides pairwise alignment, population template estimation, and alignment-induced similarity measures within a single formulation tailored to multichannel EEG.

\section{EEGSync: Complex Unitary Temporal Alignment}
\label{sec:eeg_sync}

Consider a multichannel electroencephalogram recording from subject $s$, represented as a real-valued matrix $X_s \in \mathbb{R}^{V \times T}$, where $V$ denotes the number of spatial channels and $T$ denotes the number of temporal samples. The primary objective of \mbox{EEGSync} is to align two recordings by transforming their temporal coordinates while preserving the underlying spatial electrode coordinate system. To jointly capture amplitude dynamics and quadrature phase information, \mbox{EEGSync} operates on a complex analytic signal representation.
% \subsection{Complex Analytic Representation}
% \label{subsec:analytic_rep}
Let $X_s^{(b)} \in \mathbb{R}^{V \times T}$ denote the EEG recording for subject $s$ after bandpass filtering within frequency band $b$. Its channel-wise complex analytic representation is constructed as $Z_s = X_s^{(b)} + i\,\mathcal{H}\{X_s^{(b)}\}$, where $\mathcal{H}\{\cdot\}$ denotes the Hilbert transform applied along the temporal dimension.

The analytic representation embeds the observed EEG signal together with its quadrature component into a complex vector space, defining the domain on which \mbox{EEGSync} performs temporal alignment. This formulation retains both amplitude dynamics and phase trajectories of the band-limited signal while allowing temporal alignment to be expressed naturally through complex unitary transformations.

Prior to alignment, analytic matrices may undergo temporal mean-centering or Frobenius norm normalization depending on the preprocessing configuration. When normalization is applied, all subsequent alignment operations and geometric invariance results are defined with respect to these normalized analytic representations.

\subsection{Unitary Temporal Alignment}
\label{subsec:registration}

Let $Z_A, Z_B \in \mathbb{C}^{V \times T}$ denote two complex analytic EEG recordings of equal duration, where $Z_A$ serves as the reference and $Z_B$ as the target. \mbox{EEGSync} aligns the target through a right-acting temporal transformation $Z_B \mapsto Z_BU$, where $U \in \mathbb{C}^{T \times T}$ is constrained to the unitary group $\mathcal{U}(T)$ and satisfies $U^\dagger U = I_T$. Because $U$ acts strictly from the right, it reparameterizes the temporal coordinate space without mixing the $V$ electrode channels.

The optimal transformation is obtained by solving $U^\star = \arg\min_{U^\dagger U=I_T}\|Z_A-Z_BU\|_F^2$. Expanding the objective gives $\|Z_A-Z_BU\|_F^2=\|Z_A\|_F^2+\|Z_B\|_F^2-2\operatorname{Re}\{\operatorname{tr}(U^\dagger Z_B^\dagger Z_A)\}$, where unitary invariance gives $\|Z_BU\|_F=\|Z_B\|_F$.

Since the first two terms are independent of $U$, minimizing the Alignment error is equivalent to maximizing $\operatorname{Re}\{\operatorname{tr}(U^\dagger C)\}$ subject to $U^\dagger U=I_T$, where $C=Z_B^\dagger Z_A \in \mathbb{C}^{T\times T}$. This is the complex unitary Procrustes problem.

Let the singular value decomposition of the temporal cross-product be $C=P\Sigma Q^\dagger$, where $P,Q\in\mathcal{U}(T)$ and $\Sigma$ contains the nonnegative singular values of $C$. An optimal transformation is then $U^\star=PQ^\dagger$, giving the aligned target $Z_{B\rightarrow A}=Z_BU^\star$. The complete derivation using von Neumann's trace inequality is provided in Appendix~\ref{app:procrustes_proof}.

Figure~\ref{fig:eegsync_intro} illustrates this operation on a representative EEG channel, showing the target signal before and after EEGSync alignment together with the corresponding residual relative to the reference.

When $C$ is rank deficient, the optimal temporal operator is non-unique along zero-singular-value directions. This ambiguity is restricted to the corresponding null-space components, while the Procrustes construction continues to provide an optimal aligned representation.

Beyond providing closed-form temporal alignment, the unitary constraint induces exact invariances in the channel-space geometry of the analytic EEG representation, as characterized in Section~\ref{sec:geometry_preservation}.

\begin{figure*}[t]
    \centering

    % First row
    \begin{subfigure}[t]{0.49\textwidth}
        \centering
        \includegraphics[
            width=\linewidth,
            height=0.23\textheight,
            keepaspectratio
        ]{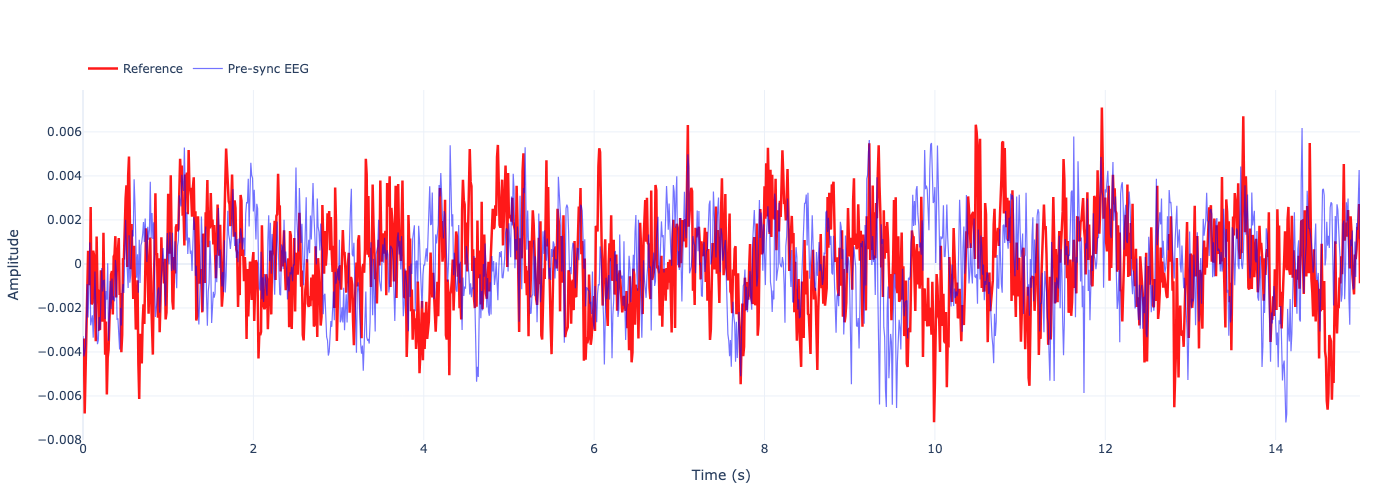}
        \caption{Reference and target EEG before alignment.}
        \label{fig:eegsync_intro_before}
    \end{subfigure}
    \hfill
    \begin{subfigure}[t]{0.49\textwidth}
        \centering
        \includegraphics[
            width=\linewidth,
            height=0.23\textheight,
            keepaspectratio
        ]{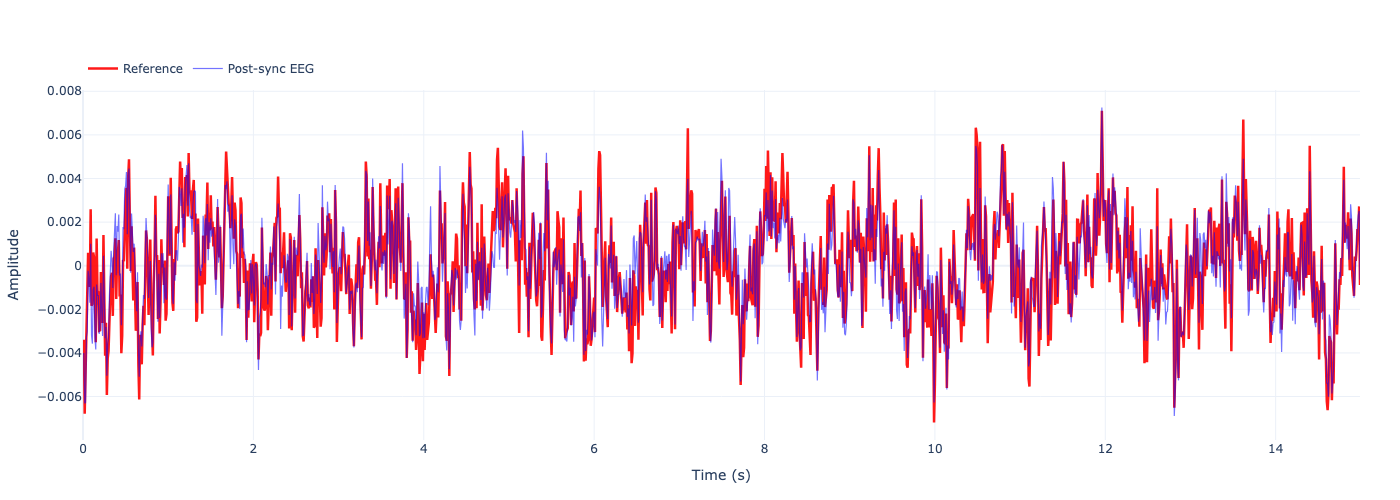}
        \caption{Reference and EEGSync-aligned target after alignment.}
        \label{fig:eegsync_intro_after}
    \end{subfigure}

    \vspace{0.5em}

    % Second row
    \begin{subfigure}[t]{0.80\textwidth}
        \centering
        \includegraphics[
            width=\linewidth,
            height=0.23\textheight,
            keepaspectratio
        ]{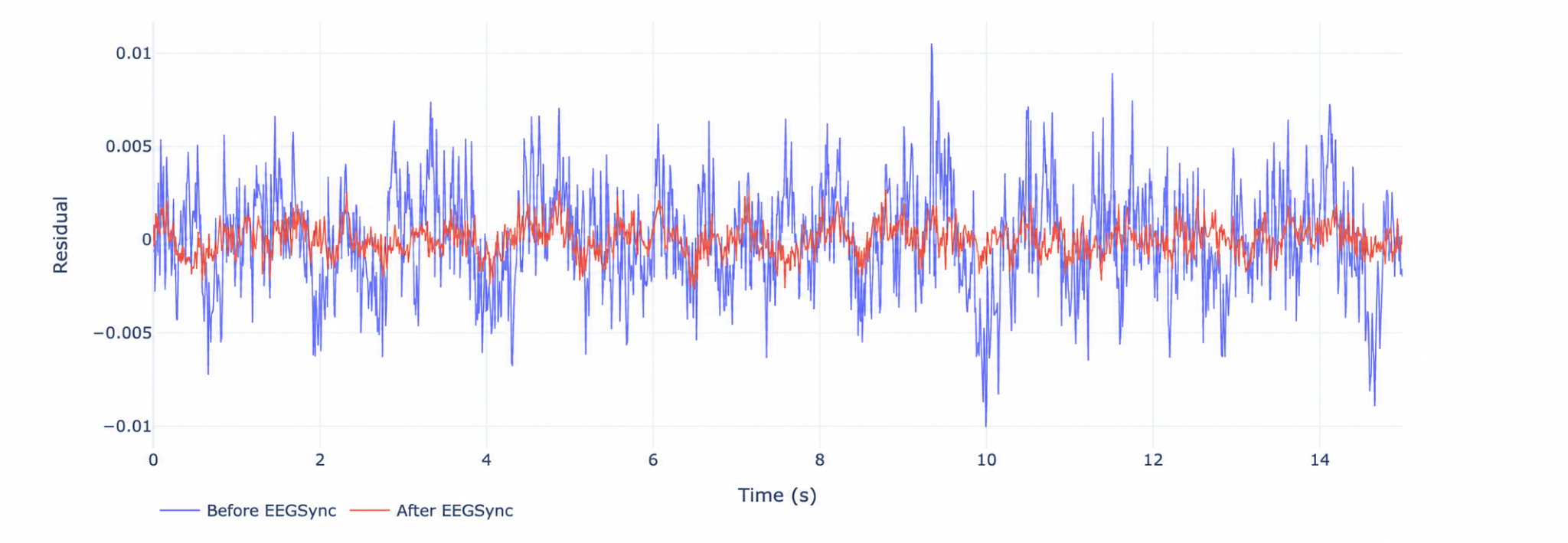}
        \caption{Residual before and after EEGSync.}
        \label{fig:eegsync_intro_residual}
    \end{subfigure}

    \caption{
    Illustrative effect of EEGSync on temporal alignment.
    (a) Reference recording and target EEG before alignment.
    (b) Reference recording and EEGSync-aligned target after alignment.
    (c) Residual between the reference and target before and after EEGSync.
    The traces correspond to a representative EEG channel from an
    intra-subject comparison, where the reference is Subject~1,
    Session~1 and the target is Subject~1, Session~2.
    EEGSync improves temporal correspondence between the target and
    reference and reduces the residual magnitude across time.
    }
    \label{fig:eegsync_intro}
\end{figure*}

\section{Geometry Preserved by EEGSync}
\label{sec:geometry_preservation}

A central property of \mbox{EEGSync} is that temporal alignment strictly preserves the spatial channel-space geometry of the target recording. While the unitary transformation reparameterizes the temporal coordinate representation, the matrix of pairwise inner products between EEG channels across electrodes remains unchanged.
\paragraph{Preservation of Channel Power and Pairwise Inner Products} 
%\label{subsec:gram_invariance}
Let $Z \in \mathbb{C}^{V \times T}$ denote an analytic EEG recording and let $\widetilde{Z}=ZU$ with $U\in\mathcal{U}(T)$ denote its transformed representation after \mbox{EEGSync} alignment. Since $UU^\dagger=I_T$, the matrix of pairwise inner products between EEG channels is preserved exactly, with $\widetilde{Z}\widetilde{Z}^{\dagger}=(ZU)(ZU)^\dagger=ZUU^\dagger Z^\dagger=ZZ^\dagger$.

The $(i,j)$-th entry of the Hermitian Gram matrix is $[ZZ^\dagger]_{ij}=Z_{i,:}Z_{j,:}^\dagger=\sum_{t=1}^{T}Z_{i,t}\overline{Z_{j,t}}$, which represents the complex inner product between the complete temporal trajectories of channels $i$ and $j$. Therefore, \mbox{EEGSync} preserves the global second-order channel-space geometry exactly. Because the diagonal entries of $ZZ^\dagger$ contain the channel energies, normalized pairwise channel correlations are also preserved when computed from consistently centered and normalized signals.
\paragraph{Preservation of Temporal Mean and Channel Correlations} 
%\label{subsec:mean_correlation_preservation}
When the analytic EEG recordings are temporally mean-centered prior to alignment, EEGSync preserves both the zero temporal mean and the exact Pearson correlation matrix of the target channels. 
Let $\mathbf{1}_T \in \mathbb{R}^T$ denote the all-ones column vector. A temporally centered target matrix $Z_B$ satisfies $Z_B \mathbf{1}_T = \mathbf{0}$. Assuming the reference $Z_A$ is also centered ($Z_A \mathbf{1}_T = \mathbf{0}$), the temporal cross-product matrix $C = Z_B^\dagger Z_A$ exhibits a strict null space property:
\begin{equation}
    C \mathbf{1}_T = Z_B^\dagger (Z_A \mathbf{1}_T) = \mathbf{0} \quad \text{and} \quad \mathbf{1}_T^T C = (Z_B \mathbf{1}_T)^\dagger Z_A = \mathbf{0}.
\end{equation}
This implies that $\mathbf{1}_T$ is simultaneously a right and left singular vector of $C$ associated with a singular value of zero. In the singular value decomposition $C = P \Sigma Q^\dagger$, the ambiguity of the unitary basis spanning the null space allows us to choose the corresponding columns of $P$ and $Q$ to be collinear with $\mathbf{1}_T$. Consequently, the optimal unitary operator $U^\star = P Q^\dagger$ satisfies $U^\star \mathbf{1}_T = \mathbf{1}_T$. Applying this temporal alignment operator to the target recording yields:
\begin{equation}
    (Z_B U^\star) \mathbf{1}_T = Z_B (U^\star \mathbf{1}_T) = Z_B \mathbf{1}_T = \mathbf{0}.
\end{equation}
Thus, the aligned representation $\tilde{Z} = Z_B U^\star$ retains a strictly zero temporal mean.

Furthermore, for temporally mean-centered analytic EEG signals,
the channel-to-channel covariance matrix is proportional to
the matrix of pairwise inner products $ZZ^\dagger$. Since
EEGSync preserves this matrix exactly, the channel-wise
variances and pairwise covariances of the complex analytic
signals remain unchanged after alignment.

Consequently, the normalized correlations between complex
EEG channels are also preserved. This guarantee applies
to the overall relationships between analytic EEG signals
and does not necessarily extend to conventional Pearson
correlations computed from the real-valued signals after
alignment.

\paragraph{Boundaries of Invariance}
%\label{subsec:invariance_boundaries}
Preservation of $ZZ^\dagger$ does not guarantee invariance of time-local or nonlinear quantities such as phase-locking value, amplitude-envelope correlation, or phase-amplitude coupling, since a general unitary transformation mixes temporal coordinates and may alter instantaneous amplitude, phase, and temporal ordering.

\section{Efficient EEGSync via Economy-SVD}
\label{sec:efficient_eeg_sync}

The full Procrustes formulation developed in Section~\ref{sec:eeg_sync} requires computing the singular value decomposition of the temporal cross-product matrix $C = Z_B^\dagger Z_A \in \mathbb{C}^{T \times T}$. In typical multichannel EEG recordings, the number of temporal samples significantly exceeds the number of spatial electrode channels ($T \gg V$). Explicitly instantiating and factorizing the $T \times T$ matrix $C$ is computationally prohibitive for long recordings, requiring $\mathcal{O}(T^2)$ memory storage.

To overcome this bottleneck, \mbox{EEGSync} admits an exact, compact formulation that shifts the dominant dense factorizations from the temporal sample dimension $T$ to the spatial channel rank $V$.

\subsection{Compact Factorization}
\label{subsec:compact_factorization}

Assume initially that $Z_A, Z_B \in \mathbb{C}^{V \times T}$ have full spatial row rank with $V \le T$. Let their economy-size singular value decompositions be $Z_A = L_A S_A R_A^\dagger$ and $Z_B = L_B S_B R_B^\dagger$, where $L_A, L_B \in \mathbb{C}^{V \times V}$ are unitary matrices, $S_A, S_B \in \mathbb{R}^{V \times V}_{>0}$ are diagonal singular value matrices, and $R_A, R_B \in \mathbb{C}^{T \times V}$ have orthonormal columns with $R_A^\dagger R_A = R_B^\dagger R_B = I_V$.

Substituting these decompositions into the temporal cross-product gives $C = Z_B^\dagger Z_A = R_B S_B L_B^\dagger L_A S_A R_A^\dagger$. Defining the compact core matrix as $M = S_B L_B^\dagger L_A S_A \in \mathbb{C}^{V \times V}$ gives the factorization $C = R_B M R_A^\dagger$.

Rather than factorizing the full $T \times T$ matrix $C$, \mbox{EEGSync} computes the SVD of the compact spatial core, $M = U_M \Sigma_M V_M^\dagger$, where $U_M, V_M \in \mathcal{U}(V)$ are $V \times V$ unitary matrices and $\Sigma_M \in \mathbb{R}^{V \times V}_{\geq 0}$ contains the singular values of $M$.

\subsection{Direct Computation of Aligned Signal} 
\label{subsec:compact_registered_rep}

In practice, downstream EEG analysis requires the aligned target recording $Z_{B \rightarrow A} = Z_B U^\star$ rather than the explicit $T \times T$ temporal operator $U^\star$. This allows direct evaluation of the aligned recording in compact form without constructing temporal null-space completions.

Using the economy-size decompositions of $Z_A$ and $Z_B$ together with the SVD of $M$, an optimal \mbox{EEGSync}-aligned signal can be written directly as $Z_{B \rightarrow A} = L_B S_B U_M V_M^\dagger R_A^\dagger$. A derivation of this expression is provided in Appendix~\ref{app:proof_economy_eegsync}. The aligned signal can therefore be constructed without explicitly instantiating either $C \in \mathbb{C}^{T \times T}$ or $U^\star \in \mathbb{C}^{T \times T}$.

The same formulation extends to rank-deficient recordings
when the reference and target have equal rank
$r = \operatorname{rank}(Z_A) = \operatorname{rank}(Z_B)$,
using reduced singular value decompositions. In this case,
the compact core becomes
$M = S_B L_B^\dagger L_A S_A \in \mathbb{C}^{r \times r}$,
and the same compact expression yields an optimal aligned
signal while preserving the channel-to-channel inner
products. When the recordings have different ranks,
additional components may be required to maintain this
preservation guarantee.

The Economy-SVD formulation also substantially reduces computational cost, requiring $\mathcal{O}(V^2T+V^3)$ time and $\mathcal{O}(VT)$ memory compared with $\mathcal{O}(T^3)$ time and $\mathcal{O}(T^2)$ memory for the full temporal Procrustes formulation. Since multichannel EEG recordings typically satisfy $T \gg V$, this provides a substantial computational advantage while preserving exact mathematical equivalence.

\section{EEGSync Induced Similarity and Distance}
\label{sec:eeg_sync_similarity_distance}

The unitary alignment framework naturally induces a similarity measure and corresponding alignment distance between two EEG recordings. Rather than introducing a separate post-alignment heuristic, the optimal value of the \mbox{EEGSync} objective itself provides a direct measure of agreement under unitary temporal reparameterization.

\paragraph{Optimal Alignment Score.}
Consider two analytic EEG matrices $Z_A, Z_B \in \mathbb{C}^{V \times T}$. \mbox{EEGSync} minimizes $\|Z_A-Z_BU\|_F^2$ over unitary transformations satisfying $U^\dagger U=I_T$. Expanding the objective gives $\|Z_A-Z_BU\|_F^2=\|Z_A\|_F^2+\|Z_B\|_F^2-2\operatorname{Re}\{\operatorname{tr}(U^\dagger Z_B^\dagger Z_A)\}$.

Let $C=Z_B^\dagger Z_A$. Under unitary temporal alignment, the largest achievable trace agreement satisfies $\max_{U^\dagger U=I_T}\operatorname{Re}\{\operatorname{tr}(U^\dagger C)\}=\|C\|_*$. Here, $\|\cdot\|_*$ denotes the nuclear norm. This quantity gives the maximum agreement attainable between the two recordings under EEGSync alignment. A derivation is provided in Appendix~\ref{app:proof_optimal_agreement}.

This leads naturally to the normalized \mbox{EEGSync} similarity score $s_{\mathrm{sync}}(Z_A,Z_B)=\|Z_B^\dagger Z_A\|_*/(\|Z_A\|_F\|Z_B\|_F)$. By von Neumann's trace inequality, the score satisfies $0\leq s_{\mathrm{sync}}(Z_A,Z_B)\leq1$. Because the nuclear norm is invariant under conjugate transposition, the similarity is also symmetric, with $s_{\mathrm{sync}}(Z_A,Z_B)=s_{\mathrm{sync}}(Z_B,Z_A)$.

A value near $1$ indicates that unitary temporal alignment can bring the two recordings into close agreement, whereas lower values indicate greater structural disagreement between their temporal trajectories. Normalized complex correlation computed after optimal \mbox{EEGSync} alignment is equal to $s_{\mathrm{sync}}(Z_A,Z_B)$ and therefore does not provide an independent post-alignment similarity measure.

\paragraph{Interpretation of EEGSync Similarity.}
Although EEGSync aligns recordings along the temporal
dimension, its similarity score reflects agreement in
the overall patterns of activity and relationships
between EEG channels, rather than whether individual
signal fluctuations occur at the same moment.
A high similarity score therefore indicates that two
recordings exhibit similar channel-level structure
under optimal temporal alignment, but does not imply
that neural events occurred at the same physical time.
This distinction is particularly important for
resting-state EEG, where recordings do not share
a common external timing reference.

\paragraph{EEGSync Alignment Distance.}
The minimum achievable alignment error defines the squared \mbox{EEGSync} distance as $d_{\mathrm{sync}}^2(Z_A,Z_B)=\min_{U^\dagger U=I_T}\|Z_A-Z_BU\|_F^2$. Using the optimal trace agreement above, this becomes $d_{\mathrm{sync}}^2(Z_A,Z_B)=\|Z_A\|_F^2+\|Z_B\|_F^2-2\|Z_B^\dagger Z_A\|_*$.

When both recordings are normalized to unit Frobenius norm, $\|Z_A\|_F=\|Z_B\|_F=1$, the expression simplifies to $d_{\mathrm{sync}}^2(Z_A,Z_B)=2\left(1-s_{\mathrm{sync}}(Z_A,Z_B)\right)$. Thus, under unit Frobenius normalization, $d_{\mathrm{sync}}$ and $s_{\mathrm{sync}}$ encode the same pairwise alignment information, with distance decreasing monotonically as similarity increases.

\section{Group EEGSync and Template Estimation}
\label{sec:group_eeg_sync}

Pairwise EEGSync aligns two recordings, whereas population analysis requires a common reference frame for a collection of recordings. We therefore extend EEGSync to group template estimation through iterative unitary temporal alignment and averaging.

\paragraph{Group Alignment Objective.}
Consider $N$ Frobenius normalized analytic EEG recordings $\mathcal{Z}=\{Z_1,\ldots,Z_N\}$ with $\|Z_i\|_F=1$. We seek a unit norm group template $T\in\mathbb{C}^{V\times T}$ together with recording specific unitary transformations $U_i\in\mathcal{U}(T)$ by minimizing $\sum_{i=1}^{N}\|T-Z_iU_i\|_F^2$ subject to $\|T\|_F=1$ and $U_i^\dagger U_i=I_T$.

Using the EEGSync alignment distance introduced in Section~\ref{sec:eeg_sync_similarity_distance}, the same objective can be written as $T^\star=\arg\min_{\|T\|_F=1}\sum_{i=1}^{N}d_{\mathrm{sync}}^2(T,Z_i)$. The resulting $T^\star$ represents a group reference after recording specific temporal variation has been accounted for.

\paragraph{Alternating Template Estimation.}
Template estimation proceeds by alternating between alignment and averaging. The initial template is obtained from the normalized mean of the unaligned recordings, $T^{(0)}=\mathcal{N}\left(\frac{1}{N}\sum_{i=1}^{N}Z_i\right)$, where $\mathcal{N}(Z)=Z/\|Z\|_F$ denotes Frobenius normalization.

At iteration $k$, each recording is independently aligned to the current template by solving $U_i^{(k)}=\arg\min_{U^\dagger U=I_T}\|T^{(k)}-Z_iU\|_F^2$, giving the aligned representation $\widetilde{Z}_i^{(k)}=Z_iU_i^{(k)}$. These alignments are computed using the Economy-SVD described in Section~\ref{sec:efficient_eeg_sync}.

The template is then updated as the normalized mean of the aligned recordings, $T^{(k+1)}=\mathcal{N}\left(\frac{1}{N}\sum_{i=1}^{N}\widetilde{Z}_i^{(k)}\right)$. The alignment step optimizes each unitary transformation for the current template, while the averaging step gives the optimal unit norm template for the current aligned recordings. The group objective therefore does not increase across iterations. In practice, we use $K=5$ refinement iterations.

\paragraph{Class Conditional EEGSync Templates.}
For supervised analysis, template estimation is performed independently within each experimental condition or diagnostic class. Let $g\in\mathcal{G}$ denote a class and let $\mathcal{S}_g^{\mathrm{train}}$ contain its training recordings. The corresponding class template $T_g=\operatorname{Template}(\mathcal{S}_g^{\mathrm{train}})$ is estimated using the same alternating alignment and averaging procedure.

For an unseen recording $Z$, agreement with class $g$
is quantified by $s_g(Z) = s_{\mathrm{sync}}(Z,T_g)$.
For binary classes $g_1$ and $g_2$, the relative
synchronization margin is defined as
$\Delta s_{\mathrm{sync}}(Z) =
s_{\mathrm{sync}}(Z,T_{g_1}) -
s_{\mathrm{sync}}(Z,T_{g_2})$.

Similarly, the relative alignment distance is defined as
\[
\Delta d_{\mathrm{sync}}(Z)
=
d_{\mathrm{sync}}(Z,T_{g_2})
-
d_{\mathrm{sync}}(Z,T_{g_1}).
\]
Both measures quantify relative agreement with the two
class templates, with positive values indicating
greater agreement with class $g_1$.

To prevent data leakage during subject independent evaluation, all class templates are estimated exclusively from training subjects within each cross validation fold. Held out participants are excluded from template initialization, iterative refinement, and feature normalization.

\paragraph{Template Interpretation.}
The estimated template $T^\star$ should be interpreted as a group representative temporal reference within the EEGSync unitary geometry rather than as a literal waveform shared across subjects. This distinction is particularly important for resting state EEG, where recordings do not share a common external timing reference. In shared stimulus settings, the template may additionally reflect temporal structure associated with the common stimulus.

\section{Experimental Framework and Method Validation}
\label{sec:experimental_framework}

We evaluate \mbox{EEGSync} from three complementary perspectives. We first verify the numerical properties, including the equivalence of the compact and full Procrustes formulations and preservation of the matrix of pairwise inner products between EEG channels. We then examine alignment behavior through intra subject and inter subject comparisons. Finally, we evaluate whether \mbox{EEGSync} representations preserve physiologically and clinically meaningful information under strict out of sample, subject independent predictive protocols.

\subsection{Datasets}
\label{subsec:datasets}

The experimental evaluation uses three EEG datasets that provide complementary validation settings.

\paragraph{Eyes Open versus Eyes Closed.}
Eyes open and eyes closed baseline recordings were obtained from the PhysioNet EEG Motor Movement and Imagery Database \citep{schalk2004bci2000,goldberger2000physiobank}. The dataset contains recordings from 109 participants acquired using the BCI2000 system with 64 scalp EEG channels arranged according to the international 10--10 electrode placement system and sampled at $160$ Hz. Each participant completed a one minute eyes open baseline recording followed by a one minute eyes closed baseline recording. These paired recordings provide a controlled within subject physiological contrast for evaluating whether \mbox{EEGSync} preserves state dependent neural structure.

\paragraph{Alzheimer's Disease, Eyes Closed.}
The OpenNeuro dataset \texttt{ds004504} contains routine eyes closed EEG recordings from 88 participants, including 36 individuals with Alzheimer's disease, 23 with frontotemporal dementia, and 29 healthy controls \citep{miltiadous2023dataset}. Recordings were acquired at the 2nd Department of Neurology, AHEPA University Hospital of Thessaloniki, using a Nihon Kohden EEG 2100 clinical EEG system. Nineteen scalp electrodes were positioned according to the international 10--20 system at Fp1, Fp2, F7, F3, Fz, F4, F8, T3, C3, Cz, C4, T4, T5, P3, Pz, P4, T6, O1, and O2, with additional mastoid electrodes A1 and A2 used for referencing. Signals were sampled at $500$ Hz with a sensitivity of $10\,\mu\mathrm{V/mm}$, a time constant of $0.3$ s, and a high frequency filter at $70$ Hz. Electrode impedance was maintained below $5\,\mathrm{k}\Omega$, and the distributed recordings used Cz as the common reference. Participants were seated with their eyes closed during acquisition. For the present binary classification analysis, participants with frontotemporal dementia were excluded, leaving the Alzheimer's disease and healthy control groups.

\paragraph{Alzheimer's Disease, Eyes Open with Photostimulation.}
OpenNeuro dataset \texttt{ds006036} provides complementary eyes open recordings from the same clinical cohort under intermittent photic stimulation \citep{ntetska2025complementary}. Recordings were acquired using the same Nihon Kohden EEG 2100 clinical EEG system with the same 19 channel international 10--20 configuration and a sampling rate of $500$ Hz. Participants remained seated with their eyes open while photic stimulation was delivered at increasing frequencies beginning at $5$ Hz and progressing in $5$ Hz increments through $10$, $15$, $20$, and, when tolerated, up to $30$ Hz. As in the eyes closed dataset, participants with frontotemporal dementia were excluded so that the analysis was restricted to Alzheimer's disease and healthy control groups.

A motor imagery experiment using the PhysioNet database is additionally included in the Appendix as a task driven validation benchmark.

\subsection{Common Signal Preprocessing}
\label{subsec:preprocessing}

All recordings were standardized using a common
processing pipeline before applying \mbox{EEGSync}.
The electrode configuration was retained separately
for each dataset, with 64 channels for the PhysioNet
eyes-open and eyes-closed recordings and 19 channels
for the Alzheimer's disease recordings. All signals
underwent zero-phase bandpass filtering between
$0.5$ and $45$ Hz, notch filtering at $50$ and $60$ Hz,
common-average referencing, resampling to
$f_s=100$ Hz, and channel-wise temporal
standardization.

Let $X_s\in\mathbb{R}^{V\times T_s}$ denote the preprocessed recording for subject $s$. Each channel was standardized according to $\widetilde{X}_{s,v,t}=(X_{s,v,t}-\mu_{s,v})/(\sigma_{s,v}+\epsilon)$, where $\mu_{s,v}$ and $\sigma_{s,v}$ denote the temporal mean and standard deviation of channel $v$, and $\epsilon=10^{-8}$ prevents numerical division by zero.

Within each dataset, recordings were cropped to a common duration $T$ determined by the shortest valid recording and divided into four equal contiguous segments, written as $X_s=[X_{s,1}\;X_{s,2}\;X_{s,3}\;X_{s,4}]$ with $X_{s,j}\in\mathbb{R}^{V\times(T/4)}$. This segmentation provides multiple observations per participant while preserving temporal continuity within each window.

Each segment $X_{s,j}$ was converted to its complex analytic representation $Z_{s,j}\in\mathbb{C}^{V\times(T/4)}$ using the Hilbert transform described in Section~\ref{sec:eeg_sync} and normalized to unit Frobenius norm as $\widehat{Z}_{s,j}=Z_{s,j}/\|Z_{s,j}\|_F$. All subsequent alignment, template estimation, and similarity computations operate on these normalized analytic matrices.

\subsection{Subject Independent Evaluation Protocol}
\label{subsec:subject_independent_eval}

All predictive experiments use subject disjoint cross validation to prevent information leakage across participants. Subjects are partitioned using five fold stratified grouped cross validation with participant identity as the grouping variable, ensuring that all segments belonging to one participant appear entirely within either the training or testing partition of a given fold.

Within each fold, class conditional templates are estimated using only training recordings, iterative template refinement does not access test data, feature scalers and normalization parameters are fitted only on the training split, and classifiers are trained without information from held out subjects. The resulting templates are then frozen and used to compute out of sample \mbox{EEGSync} similarity scores for the test participants.

\subsection{Validation Tasks}
\label{subsec:validation_tasks}

The first validation examines whether the Economy SVD implementation reproduces the direct Procrustes solution. For benchmark recordings where construction of the full $T\times T$ operator is computationally feasible, the aligned representation from the direct formulation is compared with that obtained from the compact formulation in Section~\ref{sec:efficient_eeg_sync}. Relative reconstruction error is measured as $E_{\mathrm{compact}}=\|Z_{B\rightarrow A}^{\mathrm{full}}-Z_{B\rightarrow A}^{\mathrm{compact}}\|_F/\|Z_{B\rightarrow A}^{\mathrm{full}}\|_F$.

The second validation examines whether the predicted spatial invariance is preserved numerically. For an aligned recording $\widetilde{Z}=ZU^\star$, the relative Hermitian Gram error is computed as $E_{\mathrm{Gram}}=\|(ZU^\star)(ZU^\star)^\dagger-ZZ^\dagger\|_F/\|ZZ^\dagger\|_F$. Based on the invariance established in Section~\ref{sec:geometry_preservation}, this quantity should approach zero up to floating point precision. Nonlinear quantities such as phase locking value and phase amplitude coupling are treated separately as empirical distortion diagnostics because they are not guaranteed to remain invariant.

The third validation examines intra subject and inter subject alignment behavior independently of downstream classification. Electrode wise mismatch is computed before and after alignment for recordings from the same participant and for recordings from different participants. These errors are averaged across all valid comparisons to characterize how much temporal mismatch is removed within and across subjects.

The fourth validation examines whether \mbox{EEGSync} representations retain biologically meaningful structure in unseen participants. We evaluate physiological state discrimination using eyes open versus eyes closed recordings and clinical group discrimination using Alzheimer's disease versus healthy control recordings across two complementary recording conditions from the same clinical cohort.  Segment level similarity scores $s_{\mathrm{sync}}(Z_{s,j},T_g)$ are aggregated across the four segments of each participant to obtain subject level representations. For binary problems, the relative synchronization margin $\Delta s_{\mathrm{sync}}$ is evaluated as described in Section~\ref{sec:group_eeg_sync}.

Throughout the classification experiments,
EEGSync-$\Delta$ denotes the two-feature representation
consisting of the relative alignment distance
$\Delta d_{\mathrm{sync}}$ and relative synchronization
similarity $\Delta s_{\mathrm{sync}}$, as defined in
Section~7. Both features are derived from the EEGSync
distance $d_{\mathrm{sync}}$ and similarity
$s_{\mathrm{sync}}$ introduced in Section~6 and quantify
the relative agreement of a recording with the two
class-specific templates.

EEGSync-Full consists of six features: the alignment
distances $d_{\mathrm{sync}}$ and similarities
$s_{\mathrm{sync}}$ to each of the two class templates,
together with their relative differences
$\Delta d_{\mathrm{sync}}$ and $\Delta s_{\mathrm{sync}}$.

Conventional pre-alignment descriptors consist of
14 statistical and 14 spectral features extracted
from the EEG recordings before temporal alignment.
The statistical features summarize variance,
skewness, kurtosis, line length, Hjorth mobility,
Hjorth complexity, and zero-crossing rate using
their mean and standard deviation across channels.
The spectral features include the mean and standard
deviation of log-transformed power across the delta,
theta, alpha, beta, and gamma bands, together with
four band-power ratios.

These feature sets are evaluated individually and
in combination to assess their complementary
discriminative information.

\subsection{Evaluation Metrics and Statistical Testing}
\label{subsec:evaluation_metrics}

Downstream predictive performance is reported using
the area under the receiver operating characteristic
curve (AUROC), overall accuracy, balanced accuracy,
and F1-score. Metrics are computed from pooled
out-of-sample predictions across the five test folds.

Multiple feature configurations and classifiers were evaluated across the same five subject-independent cross-validation folds, and top configurations were identified by comparing their pooled out-of-fold performance. Because model selection was performed directly on these cross-validation results, the reported peak scores should be interpreted as the best observed performance within our evaluation rather than as unbiased generalization estimates for a pre-specified model.

For the paired eyes open and eyes closed comparison, statistical significance is evaluated using paired Wilcoxon signed rank tests on subject level \mbox{EEGSync} scores. 
For the Alzheimer's disease versus healthy control comparisons, statistical significance is assessed using two-sample Mann--Whitney U tests, while Kolmogorov--Smirnov tests are additionally used to characterize differences between the empirical feature distributions.
 Numerical invariance measures and alignment errors are reported separately from predictive performance so that mathematical validation remains distinct from physiological and clinical evaluation.

\subsection{Intra and Inter Subject Comparison}
\label{subsec:intra_inter_registration}

We next examine how \mbox{EEGSync} alignment changes the mismatch between EEG recordings within and across subjects.
 The purpose of this analysis is to provide a spatial view of the alignment behavior over the electrode array while retaining the original electrode coordinate system.

\begin{figure}[t]
    \centering
    \includegraphics[width=0.5\linewidth]{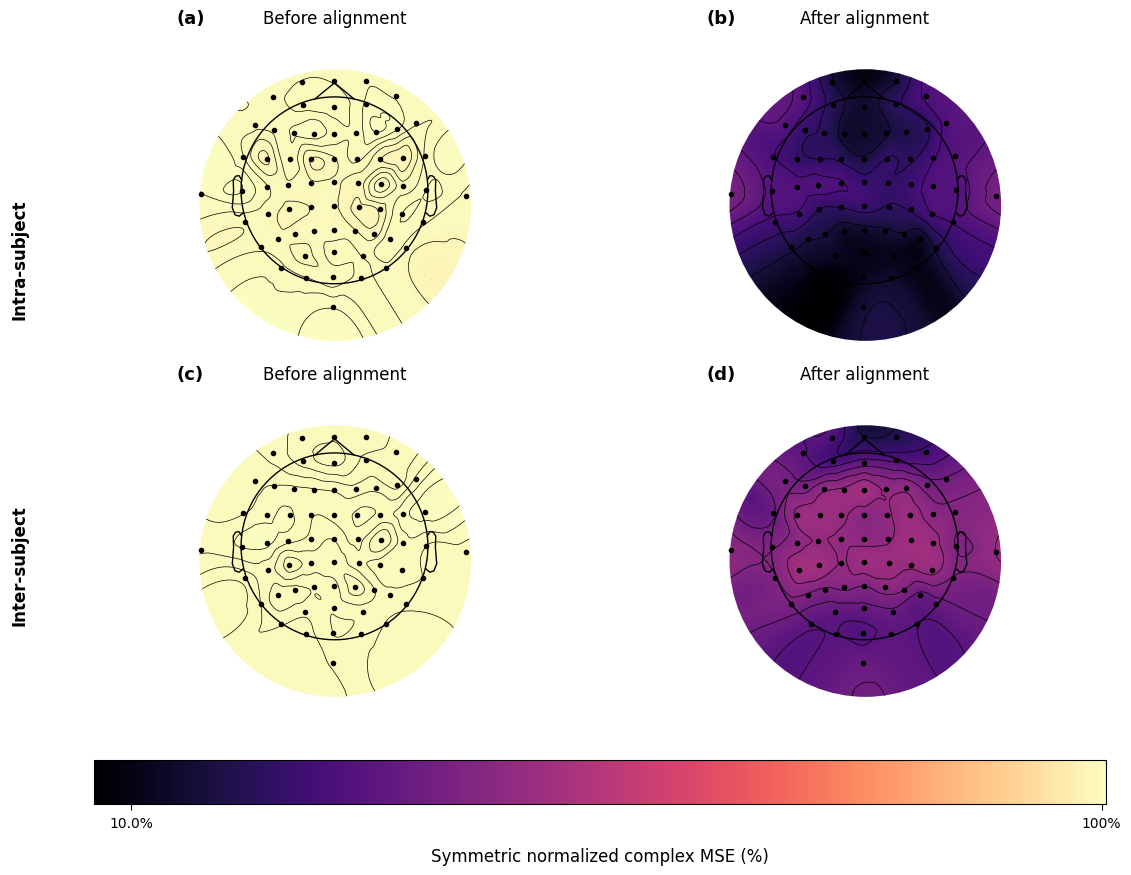}
    \caption{
    \textbf{Average intra subject and inter subject mismatch before and after EEGSync alignment.}
    Panels (a) and (b) show the mean electrode wise symmetric normalized complex MSE across intra subject comparisons before and after alignment. Panels (c) and (d) show the corresponding mean error across inter subject comparisons. The same logarithmic color scale is used for all four maps, allowing direct comparison of the spatial mismatch across the two regimes. EEGSync reduces the average mismatch in both cases, while a larger residual difference remains across subjects after alignment.
    }
    \label{fig:mse_comparison}
\end{figure}

For a reference recording $Z_A \in \mathbb{C}^{V \times T}$ and a target recording $Z_B \in \mathbb{C}^{V \times T}$, \mbox{EEGSync} produces the aligned target $Z_{B \rightarrow A}=Z_BU^\star$, where $U^\star \in \mathcal{U}(T)$. Alignment error is evaluated independently at each electrode using the complex temporal mismatch between the reference and target recordings. Before alignment, the channel wise error is $\operatorname{MSE}_v^{\mathrm{pre}}=\frac{1}{T}\sum_{t=1}^{T}|Z_{A,v,t}-Z_{B,v,t}|^2$. After alignment, the corresponding residual error is $\operatorname{MSE}_v^{\mathrm{post}}=\frac{1}{T}\sum_{t=1}^{T}|Z_{A,v,t}-Z_{B\rightarrow A,v,t}|^2$. For visualization, these errors are expressed using the symmetric normalized complex MSE percentage shown in Figure~\ref{fig:mse_comparison}.

The intra subject analysis compares recordings obtained from the same participant at different recording instances, while the inter subject analysis compares recordings from different participants acquired under the same experimental condition. For each comparison, the electrode wise mismatch is computed before and after alignment and then averaged across all valid comparisons. Figure~\ref{fig:mse_comparison} shows the resulting spatial maps, with panels (a) and (b) corresponding to the intra subject analysis and panels (c) and (d) corresponding to the inter subject analysis.

Figure~\ref{fig:mse_comparison} shows that substantial mismatch is present before alignment in both regimes. After applying \mbox{EEGSync}, the average mismatch decreases across the electrode array for both intra subject and inter subject comparisons. The intra subject recordings exhibit lower residual mismatch after alignment, while the inter subject recordings retain greater residual disagreement. This behavior is consistent with recordings from the same individual sharing more underlying structure than recordings from different individuals even after temporal coordinate alignment.

The comparison provides an intuitive view of the effect of \mbox{EEGSync} on real EEG recordings. Temporal mismatch is reduced both within and across subjects without mixing electrode channels. At the same time, alignment does not force different recordings to become identical, as greater residual mismatch remains in the inter subject setting. This reduction is achieved while preserving the matrix of pairwise inner products between EEG channels, $Z_BZ_B^\dagger$, as established in Section~\ref{sec:geometry_preservation}. The following experiments examine whether the resulting aligned representations also retain physiologically and clinically discriminative information.

\subsection{Physiological-State Validation: Eyes Open versus Eyes Closed}
\label{subsec:eo_ec_validation}

We evaluated whether \mbox{EEGSync} preserves physiological differences between resting-state eyes-open (EO) and eyes-closed (EC) conditions using the PhysioNet EEG Motor Movement/Imagery Database. Each participant contributed one EO and one EC baseline recording, with strict subject separation between training and test sets. Class-specific \mbox{EEGSync} templates were constructed using training subjects exclusively, following the standardized pipeline detailed in Section~\ref{sec:experimental_framework}.

The best subject-independent classification performance metrics are summarized in Table~\ref{tab:eo_ec_results}. The highest overall AUROC of $0.9600$ was achieved using EEGSync-$\Delta$ features combined with time-domain statistical descriptors and an $L_2$-regularized logistic regression classifier. The highest classification accuracy and balanced accuracy were both $89.45\%$, with an $F_1$-score of $0.8959$, obtained using the full \mbox{EEGSync} feature set paired with pre-alignment descriptors and a Random Forest classifier.

\begin{table}[t]
\centering
\small
\caption{\textbf{Subject-independent Eyes-Open (EO) versus Eyes-Closed (EC) classification performance.} Best-performing model configurations when optimizing for AUROC and accuracy-based metrics, respectively.}
\label{tab:eo_ec_results}
\setlength{\tabcolsep}{4pt}
\resizebox{\linewidth}{!}{%
\begin{tabular}{lllc cccc}
\toprule
\textbf{Optimized Metric} & \textbf{Feature Config.} & \textbf{Classifier} & \textbf{\# Feat.} & \textbf{AUROC} & \textbf{Acc. (\%)} & \textbf{B. Acc. (\%)} & \textbf{$F_1$} \\
\midrule
AUROC & EEGSync-$\Delta$ + Statistical & $L_2$ LogReg & 16 & \textbf{0.9600} & 86.70 & 86.70 & 0.8711 \\
\addlinespace
Acc. / B. Acc. / $F_1$ & EEGSync-Full + Pre-alignment & Random Forest & 34 & 0.9508 & \textbf{89.45} & \textbf{89.45} & \textbf{0.8959} \\
\bottomrule
\end{tabular}%
}
\end{table}

\begin{figure}[t]
\centering
\begin{subfigure}[t]{0.48\linewidth}
\centering
\includegraphics[width=\linewidth]{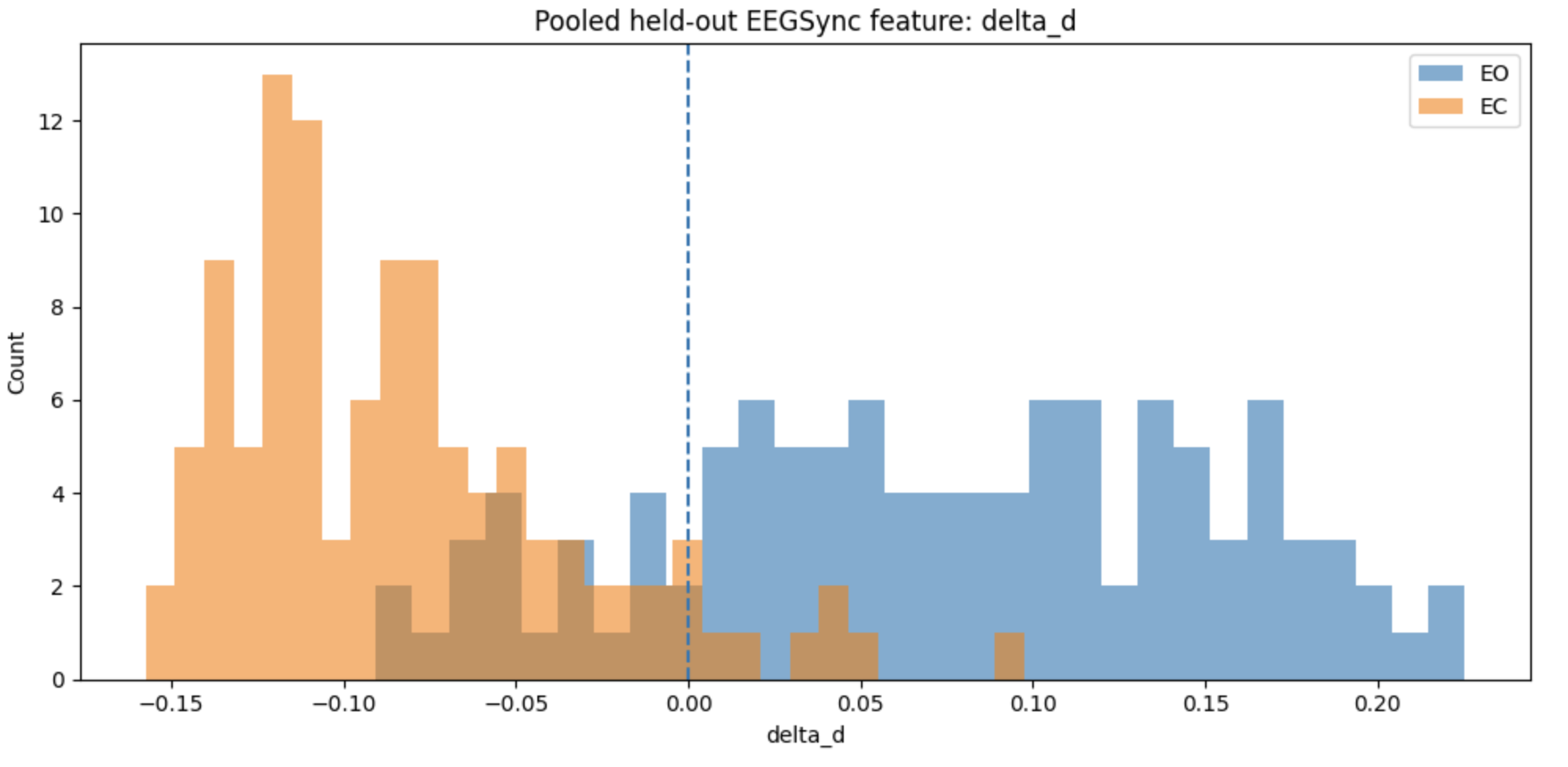}
\caption{$\Delta d_{\mathrm{sync}}$ distribution.}
\label{fig:eo_ec_dist_d}
\end{subfigure}
\hfill
\begin{subfigure}[t]{0.48\linewidth}
\centering
\includegraphics[width=\linewidth]{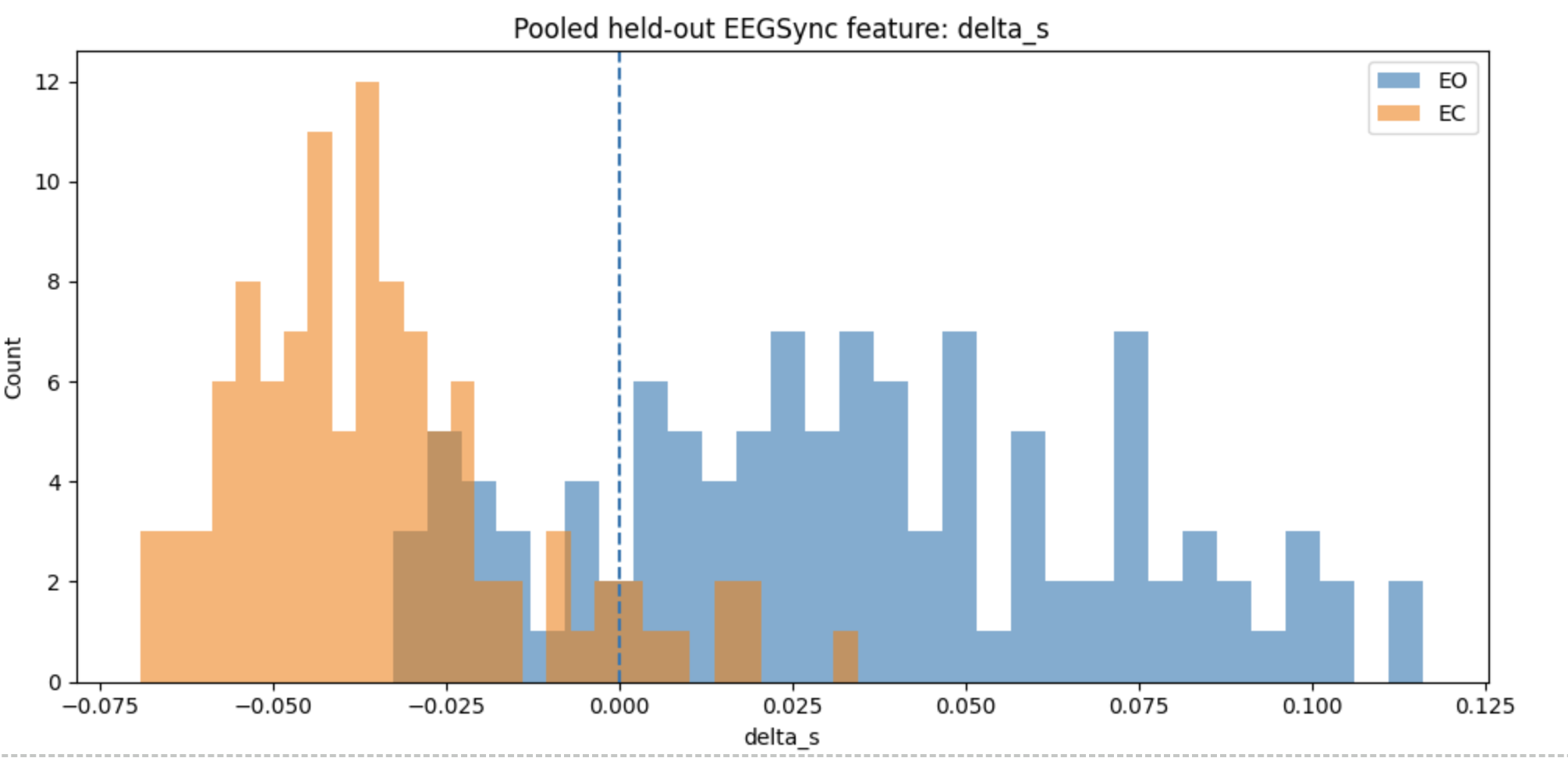}
\caption{$\Delta s_{\mathrm{sync}}$ distribution.}
\label{fig:eo_ec_dist_s}
\end{subfigure}
\caption{\textbf{Held-out EEGSync feature distributions for eyes-open and eyes-closed recordings.} Both relative alignment distance ($\Delta d_{\mathrm{sync}}$) and relative alignment similarity ($\Delta s_{\mathrm{sync}}$) exhibit clear separation between the two physiological states under strict subject-independent cross-validation.}
\label{fig:eo_ec_dist}
\end{figure}

The physiological separation is also directly visible in the empirical distributions of the held-out \mbox{EEGSync} features (Figure~\ref{fig:eo_ec_dist}). EO and EC recordings occupy substantially distinct regions of the relative synchronization space, exhibiting minimal distributional overlap around zero. When evaluated individually as univariate predictors, the single features achieved AUROCs of $0.9589$ for relative distance ($\Delta d_{\mathrm{sync}}$) and $0.9526$ for relative similarity ($\Delta s_{\mathrm{sync}}$), demonstrating that strong physiological-state information is already encoded directly within the low-dimensional \mbox{EEGSync} representation.

These empirical results confirm that \mbox{EEGSync} preserves biologically meaningful eye-state differences while yielding a compact, low-dimensional representation that generalizes reliably to previously unseen participants.

\subsection{Clinical Validation: Alzheimer's Disease versus Healthy Controls}
\label{subsec:clinical_validation_ad_hc}

We next evaluated \mbox{EEGSync} for subject-level discrimination between Alzheimer's disease (AD) patients and healthy controls (HC) using two complementary OpenNeuro datasets from the same clinical cohort: eyes-closed resting-state recordings (\texttt{ds004504}) and eyes-open recordings acquired under intermittent photic stimulation (\texttt{ds006036}). Participants with frontotemporal dementia (FTD) were excluded from the binary AD versus HC classification. Both datasets were harmonized to the standard 19-channel montage ($V = 19$) and evaluated using the strict subject-independent cross-validation protocol detailed in Section~\ref{sec:experimental_framework}.

The highest AUROC in the eyes-closed cohort reached $0.9272$, obtained by combining EEGSync-$\Delta$ features with conventional spectral power descriptors using a linear Support Vector Machine (SVM). In the complementary eyes-open photostimulation condition, EEGSync-$\Delta$ combined with spectral features achieved an AUROC of $0.8582$ and an accuracy of $78.46\%$. The corresponding balanced accuracies for these
highest-AUROC configurations were 84.38\% and
78.88\% for the eyes-closed and eyes-open cohorts,
respectively. When optimizing specifically for balanced
accuracy, the highest values reached 87.50\% and
79.21\%, respectively.

\begin{table}[t]
\centering
\small
\caption{Subject-independent Alzheimer's disease (AD) versus healthy control (HC) classification performance across two complementary recording conditions. The table summarizes the model configurations achieving the highest AUROC for the eyes-closed resting-state and eyes-open photostimulation recordings, respectively.}
\label{tab:ad_hc_results}
\setlength{\tabcolsep}{4pt}
\resizebox{\linewidth}{!}{%
\begin{tabular}{llccccc}
\toprule
\textbf{Dataset Cohort} & \textbf{Feature Config.} & \textbf{Classifier} & \textbf{AUROC} & \textbf{Acc. (\%)} & \textbf{B. Acc. (\%)} & \textbf{$F_1$} \\
\midrule
Eyes closed (\texttt{ds004504}) & EEGSync-$\Delta$ + Spectral & Linear SVM & \textbf{0.9272} & 83.08 & 84.38 & 0.8302 \\
\addlinespace
Eyes open (\texttt{ds006036}) & EEGSync-$\Delta$ + Spectral & $L_2$ LogReg & \textbf{0.8582} & 78.46 & 78.88 & 0.7941 \\
\bottomrule
\end{tabular}%
}
\end{table}

\begin{figure}[t]
\centering
\begin{subfigure}[t]{0.48\linewidth}
    \centering
    \includegraphics[width=\linewidth]{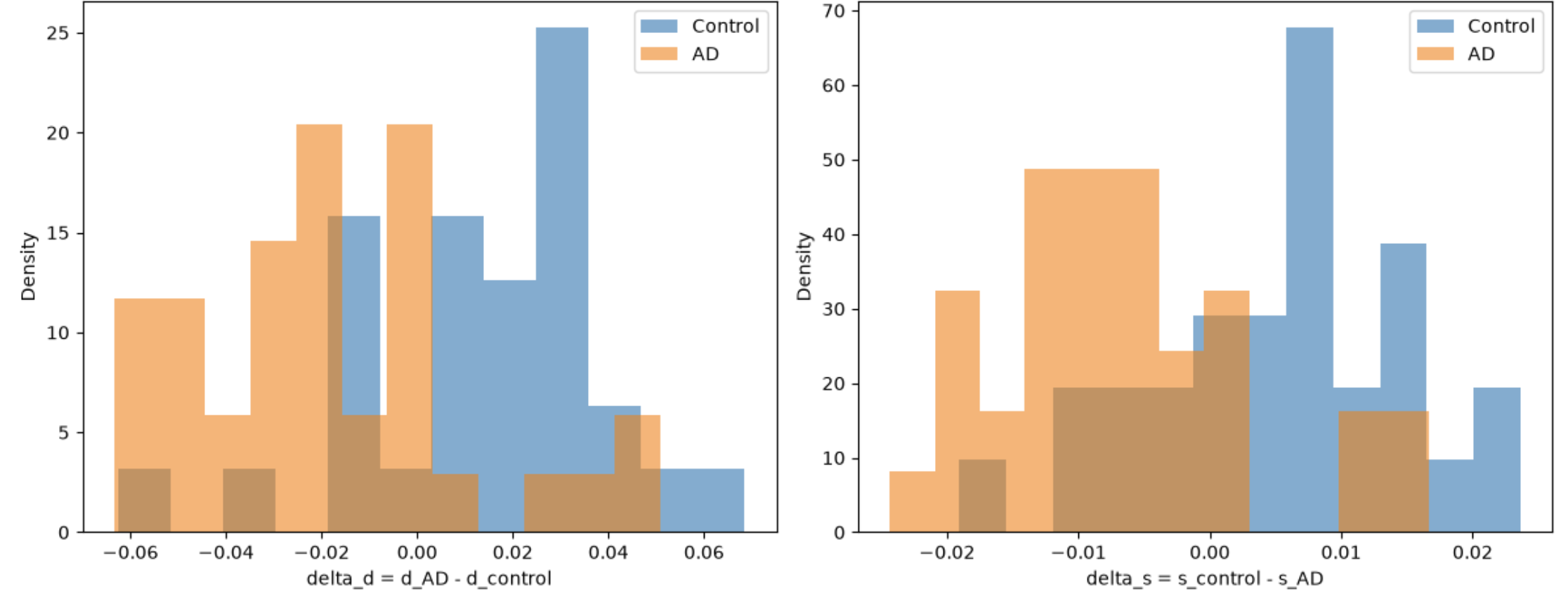}
    \caption{Eyes-closed cohort (\texttt{ds004504}).}
    \label{fig:ad_ec_dist}
\end{subfigure}
\hfill
\begin{subfigure}[t]{0.48\linewidth}
    \centering
    \includegraphics[width=\linewidth]{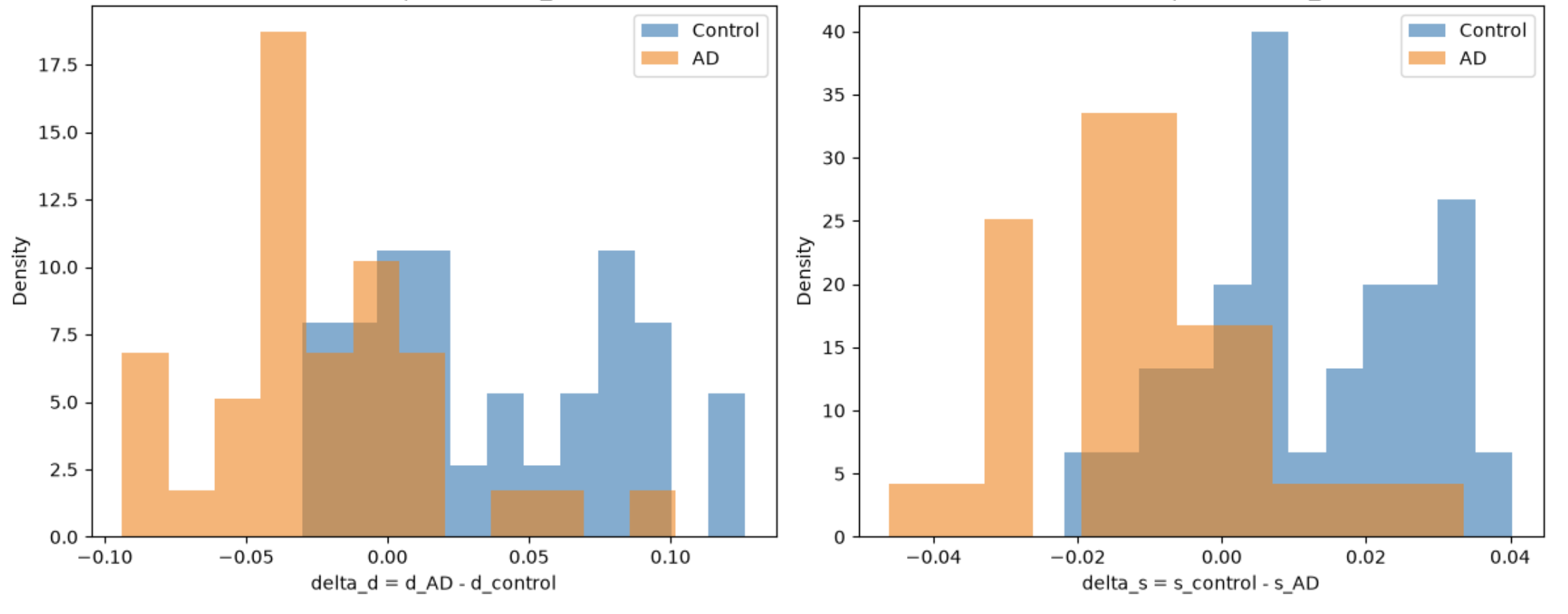}
    \caption{Eyes-open cohort (\texttt{ds006036}).}
    \label{fig:ad_eo_dist}
\end{subfigure}
\caption{\textbf{Held-out EEGSync synchronization feature distributions for Alzheimer's disease (AD) and healthy control (HC) participants.} \textbf{(a)} Eyes-closed resting-state recordings. \textbf{(b)} Eyes-open photostimulation recordings. Both cohorts exhibit disease-related shifts in the synchronization margins ($\Delta d_{\mathrm{sync}}$ and $\Delta s_{\mathrm{sync}}$), with sharper group separation observed under the eyes-closed condition.}
\label{fig:ad_distributions}
\end{figure}

Disease-associated group separation is also directly observable in the empirical distributions of held-out synchronization features (Figure~\ref{fig:ad_distributions}). In the eyes-closed cohort, healthy control subjects concentrate predominantly within positive relative synchronization values, whereas AD participants exhibit a systematic shift toward negative values. A similar separation pattern is observed in the eyes-open photostimulation recordings, although with greater group overlap than in the eyes-closed condition.

The standalone synchronization features were also discriminative without multivariate downstream models. In the eyes-closed cohort, relative distance ($\Delta d_{\mathrm{sync}}$) and relative similarity ($\Delta s_{\mathrm{sync}}$) achieved univariate AUROCs of $0.8113$ and $0.8094$, respectively. In the eyes-open cohort, the corresponding standalone AUROCs evaluated to $0.8592$ and $0.8429$. Across both recording conditions, these results indicate that disease-relevant neurophysiological structure is captured within the low-dimensional EEGSync representation under both eyes-closed resting-state and eyes-open photostimulation conditions.

\section{Discussion}
\label{sec:discussion}

\mbox{EEGSync} introduces a different perspective on temporal alignment for multichannel EEG. Rather than attempting to match individual time points through local shifts or nonlinear warping, the method aligns recordings through a unitary transformation of their temporal coordinates. Because this transformation acts exclusively along the temporal dimension, the electrode channels are not mixed. This separation between temporal alignment and electrode-space structure is central to the formulation.

An important consequence of the unitary constraint is that the channel-to-channel inner-product matrix $ZZ^\dagger$ is preserved exactly. The singular-value spectrum of the recording is preserved as well. EEGSync can consequently alter how temporal activity is represented while maintaining these global properties of the original multichannel signal. This distinguishes the method from approaches that achieve cross-subject correspondence by transforming or mixing the spatial channel representation.

The optimal unitary transformation should be understood as alignment of temporal coordinates rather than as estimation of a simple physical delay. A general unitary transformation can capture correspondences that extend beyond time shifts or local temporal deformations. 
EEGSync therefore provides a latent temporal coordinate system in which two recordings can be compared after reducing differences in their temporal organization that arise from variations across recording sessions and participants.
 This interpretation is particularly relevant for resting-state EEG, where there is no common external timeline against which participants can be directly synchronized.

The similarity and distance induced by the alignment objective provide a natural way to quantify this correspondence. The similarity score $s_{\mathrm{sync}}$ measures the maximum agreement attainable between two recordings under unitary temporal alignment, while $d_{\mathrm{sync}}$ measures the corresponding minimum residual mismatch. EEGSync uses the value of its own optimization objective to quantify how well two recordings can be brought into correspondence.

The group-template formulation extends this idea from pairwise alignment to population analysis. Instead of choosing an arbitrary participant as a reference, recordings are iteratively aligned and averaged to obtain a representative temporal reference for the group. Class-specific templates can then be estimated from training subjects and used to characterize unseen recordings through their relative agreement with each class. 

The resulting template should not be interpreted as a literal waveform shared by all participants, but rather as a common reference that accounts for differences in temporal organization across recordings and participants.

The experimental results support this interpretation of the framework. In the intra-subject and inter-subject comparisons, temporal mismatch decreased after alignment in both settings, while residual mismatch remained greater across subjects than within subjects. EEGSync reduced temporal variability within and across participants while preserving differences between recordings that could not be resolved through optimal alignment.

The Eyes-Open versus Eyes-Closed experiment further showed that the aligned representations retained physiologically meaningful information. Under strict subject-independent evaluation, the best EEGSync-based model achieved an AUROC of $0.9600$. The univariate distance and similarity margins were also strongly discriminative, with AUROCs of $0.9589$ and $0.9526$, respectively. 

Clinical validation showed a similar pattern in Alzheimer's disease classification. For the eyes-closed recordings, combining EEGSync features with spectral features achieved an AUROC of $0.9272$. The spectral-only model obtained slightly higher classification accuracy in this condition, suggesting that EEGSync features provide complementary information rather than uniformly replacing conventional spectral features. In the complementary eyes-open photostimulation recordings from the same clinical cohort, EEGSync-based models achieved AUROCs up to $0.8582$. The corresponding distance and similarity margins also showed substantial discrimination between Alzheimer's disease and healthy controls.

These findings demonstrate that EEGSync reduces temporal variability across recordings while preserving information relevant to physiological state and clinical diagnosis. EEGSync features also distinguish between physiological and clinical conditions in previously unseen participants, with all class templates estimated exclusively from training subjects within each cross-validation fold. EEGSync thus provides a framework for temporal alignment and comparison of individual recordings against population references while preserving the original electrode-channel relationships.

\section{Limitations}
\label{sec:limitations}

The flexibility of the unitary transformation introduces an important limitation in interpretation. A general element of $\mathcal{U}(T)$ is substantially broader than a physical time shift or local temporal warp. The resulting transformation should therefore be interpreted as alignment in a latent temporal coordinate system rather than as recovery of a physiological or neural clock offset. This distinction is particularly important in resting-state EEG, where no shared external timing reference exists.

The exact preservation guarantee is also limited to quantities encoded by $ZZ^\dagger$. Measures that depend on local temporal structure or nonlinear relationships, including phase-locking value, amplitude-envelope correlation, and phase--amplitude coupling, are not generally invariant under an unrestricted unitary transformation. Changes in these quantities should therefore be treated as diagnostics of alignment-induced distortion rather than as theoretically preserved properties of \mbox{EEGSync}.

The present clinical evaluation uses two complementary recording conditions from the same underlying clinical cohort rather than independent external cohorts. Validation across additional institutions, acquisition systems, and patient populations will therefore be necessary to establish broader clinical generalizability. Similarly, the current physiological experiments focus primarily on resting-state conditions, while the behavior of \mbox{EEGSync} under naturalistic continuous stimulation remains to be evaluated.

Finally, the current formulation places no explicit locality, smoothness, or frequency-specific constraints on the unitary transformation. Future extensions could introduce such structure to provide a continuum between unrestricted latent temporal alignment and more physically interpretable temporal transformations. Naturalistic shared-stimulus paradigms may be particularly useful for studying this trade-off because the common stimulus provides an external temporal reference against which the optimal unitary alignment can be evaluated.

\section{Conclusion}
\label{sec:conclusion}

We present \mbox{EEGSync}, a complex unitary temporal alignment framework for multichannel EEG recordings. By applying unitary transformations exclusively along the temporal dimension, \mbox{EEGSync} reduces temporal mismatch without mixing electrode channels and exactly preserves the channel-to-channel inner-product matrix $ZZ^\dagger$. The Economy-SVD formulation reproduces the full temporal Procrustes solution to machine precision while avoiding construction of the $T\times T$ temporal operator, making the method practical for long recordings where $T\gg V$.

Empirical validation confirmed both the mathematical and practical behavior of the framework. The compact and full formulations agreed to approximately machine precision, while the channel inner-product matrix and singular spectrum remained unchanged after alignment. \mbox{EEGSync} also reduced temporal mismatch in both intra-subject and inter-subject comparisons without forcing distinct recordings to become identical.

The aligned representations retained physiologically and clinically discriminative information. \mbox{EEGSync}-based features achieved a subject-independent AUROC of $0.9600$ for Eyes-Open versus Eyes-Closed classification and up to $0.9272$ for Alzheimer's disease identification in the eyes-closed recordings, with additional discrimination observed under the complementary eyes-open photostimulation condition.

\mbox{EEGSync} provides a unified framework for pairwise temporal alignment, population template estimation, and comparison of recordings through alignment-induced similarity and distance. Future extensions incorporating structured temporal constraints and evaluation in naturalistic shared-stimulus settings may further improve the interpretability and applicability of the framework.

\section*{Author Contributions}

AJ: concepts, derivation
SB: experiments, writing
RL: guidance, funding, mentoring, proofreading

\section*{Funding}

This work is supported by NIH grants R01EB026299, R01NS121761 and DoD grant HT94252310149.

\section*{Declaration of Competing Interests}

None.

\printbibliography

\appendix

\section{Appendix}
\label{sec:appendix}

\subsection{Derivation of the Optimal Unitary Alignment}
\label{app:procrustes_proof}

The EEGSync alignment objective minimizes $\|Z_A-Z_BU\|_F^2$ over unitary transformations satisfying $U^\dagger U=I_T$. Since $\|Z_A\|_F^2$ and $\|Z_B\|_F^2$ are independent of $U$, the optimization is equivalent to maximizing $\operatorname{Re}\{\operatorname{tr}(U^\dagger C)\}$, where $C=Z_B^\dagger Z_A$.

Let $C=P\Sigma Q^\dagger$ denote the singular value decomposition of the temporal cross-product. Substitution into the trace objective gives
\begin{equation}
\operatorname{Re}\left\{
\operatorname{tr}\left(
U^\dagger P\Sigma Q^\dagger
\right)
\right\}
=
\operatorname{Re}\left\{
\operatorname{tr}\left(
Q^\dagger U^\dagger P\Sigma
\right)
\right\}.
\end{equation}

Defining $W=P^\dagger UQ$, which is also unitary, the objective becomes $\operatorname{Re}\{\operatorname{tr}(W^\dagger\Sigma)\}$. By von Neumann's trace inequality,
\begin{equation}
\operatorname{Re}\left\{
\operatorname{tr}\left(
W^\dagger\Sigma
\right)
\right\}
\leq
\sum_{k=1}^{T}\sigma_k(C).
\end{equation}

For full-rank $C$, equality is attained when $W=I_T$, which gives $U^\star=PQ^\dagger$. When $C$ is rank deficient, the optimizer is non-unique along the zero singular-value subspace, while $U^\star=PQ^\dagger$ remains a valid optimal solution. The aligned target is therefore given by $Z_{B\rightarrow A}=Z_BU^\star$.

\subsection{Derivation of the Economy-SVD Alignment}
\label{app:proof_economy_eegsync}

Let the economy-size singular value decompositions of the reference and target recordings be $Z_A=L_AS_AR_A^\dagger$ and $Z_B=L_BS_BR_B^\dagger$. The temporal cross-product can then be written as
\begin{equation}
C
=
Z_B^\dagger Z_A
=
R_BS_BL_B^\dagger L_AS_AR_A^\dagger.
\end{equation}

Defining the compact core matrix $M=S_BL_B^\dagger L_AS_A$ gives $C=R_BMR_A^\dagger$. Let the singular value decomposition of the compact core be $M=U_M\Sigma_MV_M^\dagger$. Substitution yields
\begin{equation}
C
=
R_BU_M\Sigma_MV_M^\dagger R_A^\dagger.
\end{equation}

Because $R_A$ and $R_B$ have orthonormal columns, this factorization identifies the nonzero singular subspaces of the full temporal cross-product. Under the full-row-rank assumption, a complete unitary optimizer can be constructed as
\begin{equation}
U^\star
=
R_BU_MV_M^\dagger R_A^\dagger
+
R_B^\perp W(R_A^\perp)^\dagger,
\end{equation}
where $R_A^\perp$ and $R_B^\perp$ are orthonormal bases for the complements of the column spaces of $R_A$ and $R_B$, respectively, and $W\in\mathcal{U}(T-V)$ is arbitrary. The second term completes the transformation to a full unitary operator and vanishes after multiplication by $Z_B$, since $Z_BR_B^\perp=0$.

Using $Z_B=L_BS_BR_B^\dagger$ and $R_B^\dagger R_B=I_V$,
\begin{align}
Z_BU^\star
&=
\left(L_BS_BR_B^\dagger\right)
\left(
R_BU_MV_M^\dagger R_A^\dagger+U_\perp
\right)
\nonumber\\
&=
L_BS_BU_MV_M^\dagger R_A^\dagger.
\end{align}

Hence the aligned target can be constructed directly as
\begin{equation}
Z_{B\rightarrow A}
=
L_BS_BU_MV_M^\dagger R_A^\dagger,
\end{equation}
without explicitly constructing either the $T\times T$ temporal cross-product or the full unitary transformation.

\subsection{Derivation of the Optimal EEGSync Agreement}
\label{app:proof_optimal_agreement}

Let $C=Z_B^\dagger Z_A=P\Sigma Q^\dagger$ denote the singular value decomposition of the temporal cross-product. From the optimal unitary alignment derived in Appendix~\ref{app:procrustes_proof}, one optimal transformation is $U^\star=PQ^\dagger$. Substituting this transformation into the trace objective gives
\begin{align}
\operatorname{Re}\left\{
\operatorname{tr}\left(
U^{\star\dagger}C
\right)
\right\}
&=
\operatorname{Re}\left\{
\operatorname{tr}\left(
QP^\dagger P\Sigma Q^\dagger
\right)
\right\}
\nonumber\\
&=
\operatorname{tr}(\Sigma)
\nonumber\\
&=
\sum_{k=1}^{T}\sigma_k(C)
\nonumber\\
&=
\|C\|_*.
\end{align}

Therefore, the maximum agreement attainable under unitary temporal alignment is
\begin{equation}
\max_{U^\dagger U=I_T}
\operatorname{Re}\left\{
\operatorname{tr}\left(
U^\dagger Z_B^\dagger Z_A
\right)
\right\}
=
\left\|Z_B^\dagger Z_A\right\|_*.
\end{equation}

This nuclear-norm agreement is the quantity used to define the normalized EEGSync similarity score in the main text.

% ============================================================
\subsection{Additional Eyes-Open versus Eyes-Closed Results}
\label{app:eo_ec}
% ============================================================

\begin{figure}[htbp]
    \centering

    \begin{subfigure}[t]{0.48\linewidth}
        \centering
        \includegraphics[width=\linewidth]{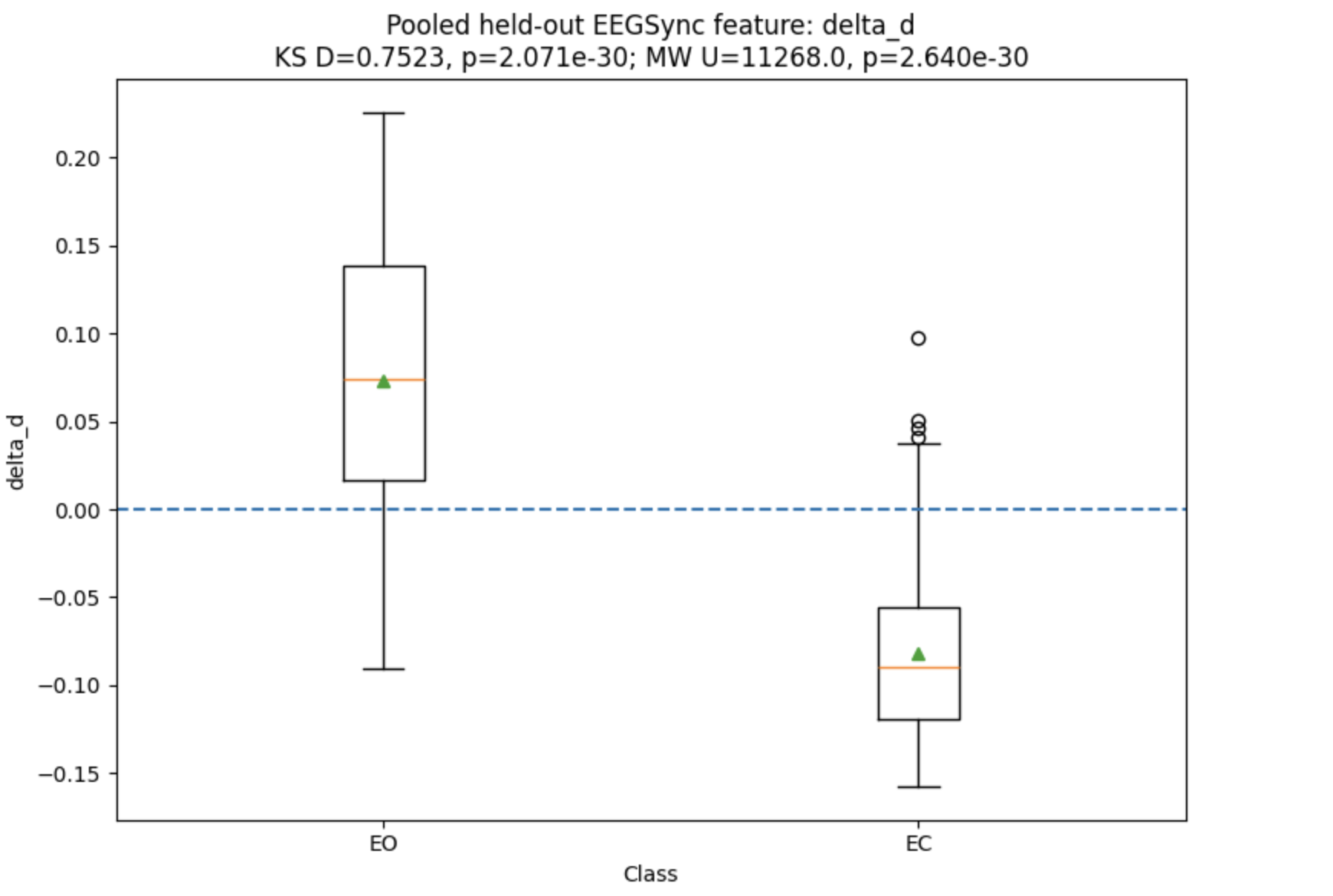}
        \caption{$\Delta d$.}
        \label{fig:app_eo_ec_d}
    \end{subfigure}
    \hfill
    \begin{subfigure}[t]{0.48\linewidth}
        \centering
        \includegraphics[width=\linewidth]{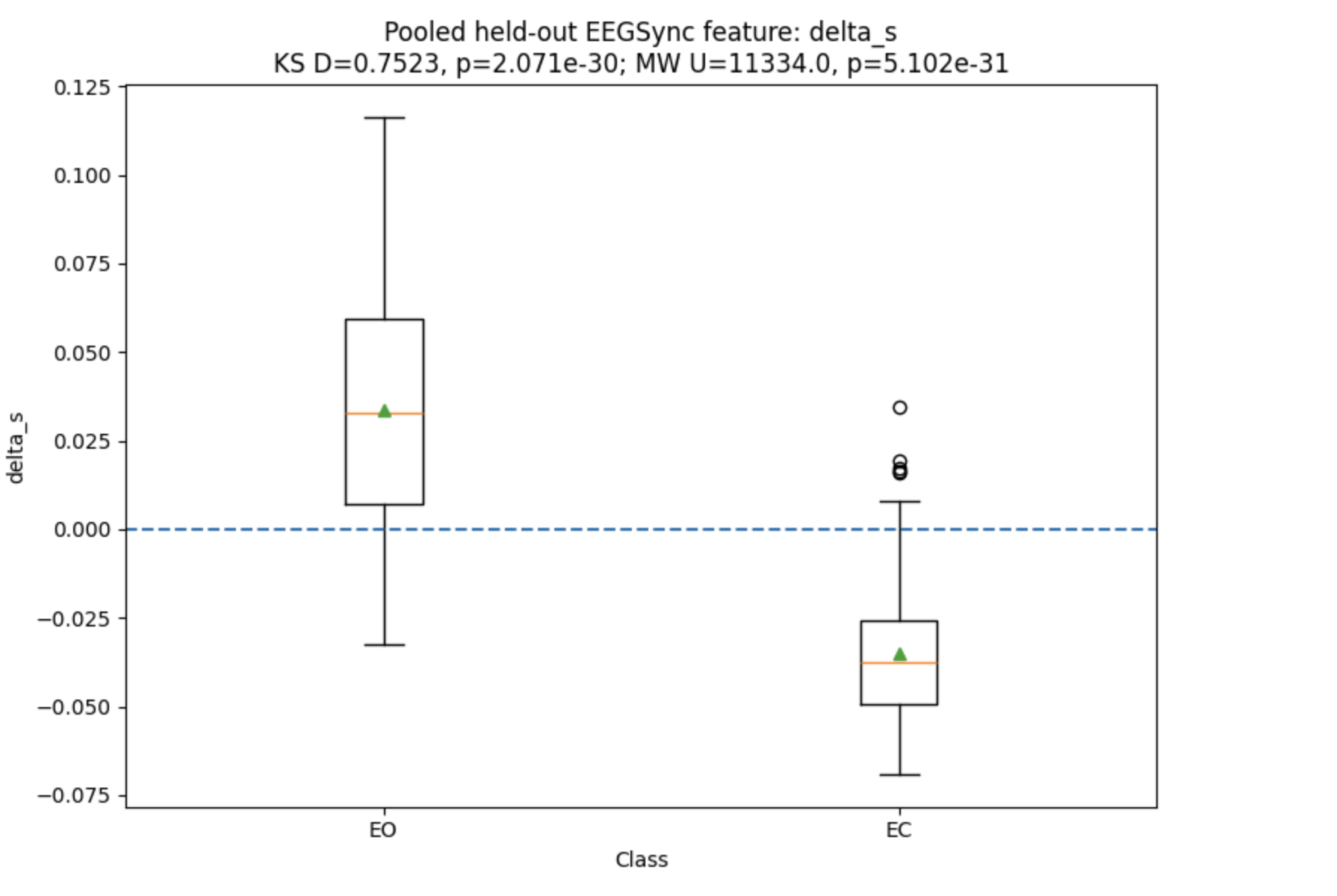}
        \caption{$\Delta s$.}
        \label{fig:app_eo_ec_s}
    \end{subfigure}

    \caption{
    \textbf{Additional EO versus EC synchronization-feature summaries.}
    Boxplots of the class-relative alignment distance $\Delta d$ and similarity $\Delta s$ across held-out recordings under subject-independent cross-validation.
    }
    \label{fig:app_eo_ec_boxplots}
\end{figure}

The main text emphasizes the distributional separation of the held-out eyes-open (EO) and eyes-closed (EC) synchronization features. Figure~\ref{fig:app_eo_ec_boxplots} provides a complementary summary using boxplots. The shifts in medians and interquartile ranges are consistent with the histogram separation reported in the main text.

% ============================================================
\subsection{Detailed Alzheimer's Disease Results}
\label{app:ad}
% ============================================================

The main text reports the highest-AUROC configuration for each Alzheimer's disease cohort. Here we provide additional model configurations and feature summaries for the eyes-closed and eyes-open datasets.

% ------------------------------------------------------------
\subsubsection{Eyes-Closed Cohort (\texttt{ds004504})}
\label{app:ad_ec}
% ------------------------------------------------------------

For the eyes-closed cohort, EEGSync-$\Delta$ combined with spectral descriptors produced the highest AUROC of $0.9272$. Spectral descriptors alone produced the highest accuracy and balanced accuracy, while EEGSync-$\Delta$ combined with spectral features achieved the highest $F_1$ score when paired with a Random Forest classifier.

\begin{table}[htbp]
\centering
\small
\caption{
Detailed AD versus HC classification results for the eyes-closed cohort (\texttt{ds004504}).
}
\label{tab:app_ad_ec}
\setlength{\tabcolsep}{3pt}
\resizebox{\linewidth}{!}{%
\begin{tabular}{lllccccc}
\toprule
\textbf{Optimized Metric}
& \textbf{Feature Config.}
& \textbf{Classifier}
& \textbf{Dim.}
& \textbf{AUROC}
& \textbf{Acc.}
& \textbf{B. Acc.}
& \textbf{$F_1$}
\\
\midrule

AUROC
& EEGSync-$\Delta$ + Spectral
& Linear SVM
& 16
& \textbf{0.9272}
& 83.08\%
& 84.38\%
& 0.8302
\\

\addlinespace

Acc. / B. Acc.
& Spectral
& Linear SVM
& 14
& 0.9148
& \textbf{86.15\%}
& \textbf{87.50\%}
& 0.8615
\\

\addlinespace

$F_1$-score
& EEGSync-$\Delta$ + Spectral
& Random Forest
& 16
& 0.9213
& 86.15\%
& 86.88\%
& \textbf{0.8696}
\\

\bottomrule
\end{tabular}%
}
\end{table}

\begin{figure}[htbp]
    \centering
    \includegraphics[width=0.78\linewidth]{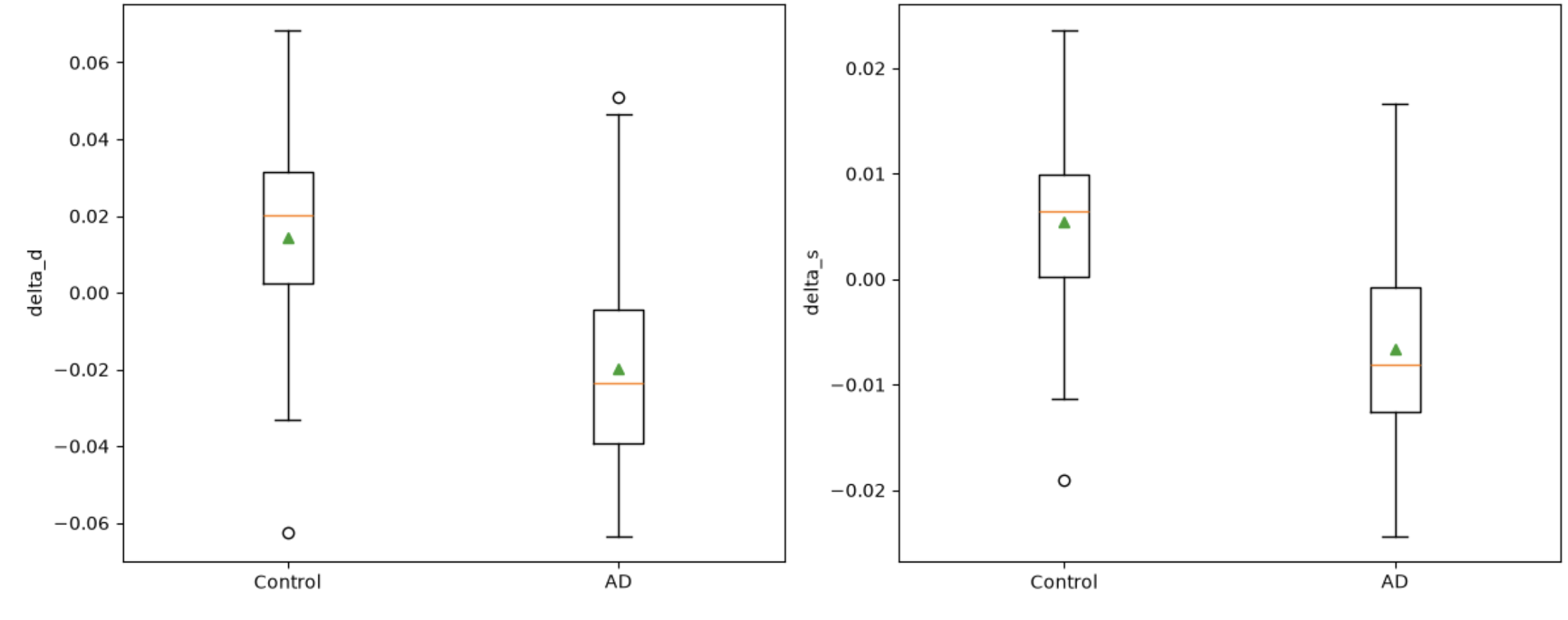}
    \caption{
    \textbf{Eyes-closed Alzheimer's disease cohort.}
    Boxplots of $\Delta d$ and $\Delta s$ for Alzheimer's disease and healthy-control subjects.
    }
    \label{fig:app_ad_ec_box}
\end{figure}

When considered individually, the synchronization features achieved AUROCs of $0.8113$ for $\Delta d$ and $0.8094$ for $\Delta s$.

\begin{table}[htbp]
\centering
\small
\caption{
Univariate synchronization-feature analysis for the eyes-closed Alzheimer's disease cohort.
}
\label{tab:app_ad_ec_uni}
\setlength{\tabcolsep}{6pt}
\begin{tabular}{lccccc}
\toprule
\textbf{Feature}
& \textbf{AUROC}
& \textbf{KS Stat.}
& \textbf{KS $p$}
& \textbf{MWU}
& \textbf{MWU $p$}
\\
\midrule

$\Delta d$
& \textbf{0.8113}
& 0.6197
& $2.44\times10^{-6}$
& 847
& $1.85\times10^{-5}$
\\

$\Delta s$
& \textbf{0.8094}
& 0.5642
& $2.89\times10^{-5}$
& 845
& $2.08\times10^{-5}$
\\

\bottomrule
\end{tabular}
\end{table}

% ------------------------------------------------------------
\subsubsection{Eyes-Open Cohort (\texttt{ds006036})}
\label{app:ad_eo}
% ------------------------------------------------------------

For the eyes-open photostimulation recordings,
EEGSync-$\Delta$ combined with spectral descriptors
achieved the highest AUROC among the evaluated
multivariate classifiers ($0.8582$), with an accuracy
of $78.46\%$. The standalone relative alignment
distance feature achieved a slightly higher AUROC
of $0.8592$. EEGSync-$\Delta$ alone produced the
highest balanced accuracy, while EEGSync-$\Delta$
combined with spectral descriptors and Gaussian
Na\"ive Bayes produced the highest $F_1$ score.

\begin{table}[htbp]
\centering
\small
\caption{
Detailed AD versus HC classification results for the eyes-open cohort (\texttt{ds006036}).
}
\label{tab:app_ad_eo}
\setlength{\tabcolsep}{3pt}
\resizebox{\linewidth}{!}{%
\begin{tabular}{lllccccc}
\toprule
\textbf{Optimized Metric}
& \textbf{Feature Config.}
& \textbf{Classifier}
& \textbf{Dim.}
& \textbf{AUROC}
& \textbf{Acc.}
& \textbf{B. Acc.}
& \textbf{$F_1$}
\\
\midrule

AUROC / Acc.
& EEGSync-$\Delta$ + Spectral
& $L_2$ LogReg
& 16
& \textbf{0.8582}
& \textbf{78.46\%}
& 78.88\%
& 0.7941
\\

\addlinespace

B. Accuracy
& EEGSync-$\Delta$
& Random Forest
& 2
& 0.8171
& 78.46\%
& \textbf{79.21\%}
& 0.7879
\\

\addlinespace

$F_1$-score
& EEGSync-$\Delta$ + Spectral
& Gaussian NB
& 16
& 0.8463
& 78.46\%
& 78.21\%
& \textbf{0.8056}
\\

\bottomrule
\end{tabular}%
}
\end{table}

\begin{figure}[htbp]
    \centering
    \includegraphics[width=0.78\linewidth]{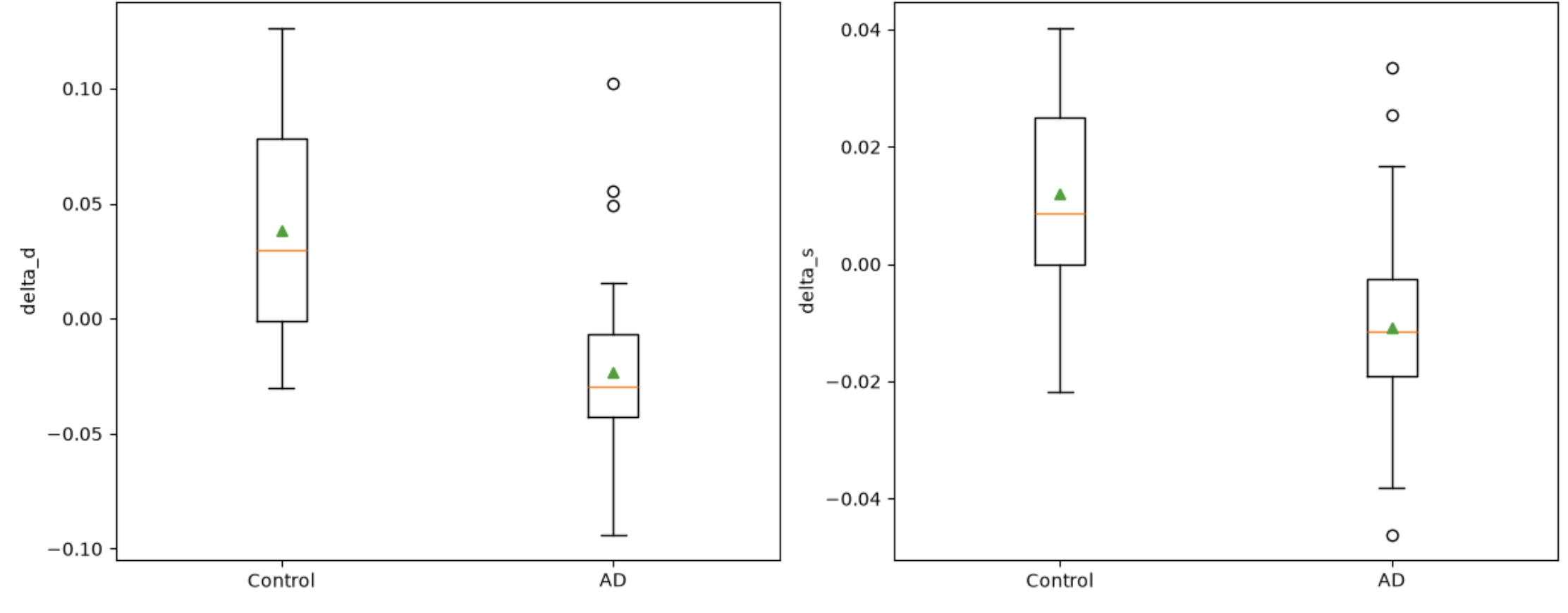}
    \caption{
    \textbf{Eyes-open Alzheimer's disease cohort.}
    Boxplots of $\Delta d$ and $\Delta s$ for Alzheimer's disease and healthy-control subjects.
    }
    \label{fig:app_ad_eo_box}
\end{figure}

The individual synchronization features achieved AUROCs of $0.8592$ for $\Delta d$ and $0.8429$ for $\Delta s$.

\begin{table}[htbp]
\centering
\small
\caption{
Univariate synchronization-feature analysis for the eyes-open Alzheimer's disease cohort.
}
\label{tab:app_ad_eo_uni}
\setlength{\tabcolsep}{6pt}
\begin{tabular}{lccccc}
\toprule
\textbf{Feature}
& \textbf{AUROC}
& \textbf{KS Stat.}
& \textbf{KS $p$}
& \textbf{MWU}
& \textbf{MWU $p$}
\\
\midrule

$\Delta d$
& \textbf{0.8592}
& 0.5843
& $1.31\times10^{-5}$
& 147
& $7.72\times10^{-7}$
\\

$\Delta s$
& \textbf{0.8429}
& 0.5776
& $1.73\times10^{-5}$
& 164
& $2.38\times10^{-6}$
\\

\bottomrule
\end{tabular}
\end{table}

% ============================================================
\subsection{Motor-Imagery Stress Test}
\label{app:motor_imagery}
% ============================================================

As an additional stress test, we evaluated EEGSync on left-hand (LH) versus right-hand (RH) motor imagery using the PhysioNet EEG Motor Movement/Imagery database. Unlike the resting-state eye-condition and Alzheimer's disease experiments, motor imagery involves more localized task-specific differences and therefore provides a substantially more challenging cross-subject discrimination problem. All preprocessing, template construction, feature extraction, classifier training, and evaluation followed the common subject-grouped pipeline described in Section~\ref{sec:experimental_framework}.

The highest AUROC was $0.7150$, obtained using EEGSync-$\Delta$ features combined with statistical descriptors and a Ridge classifier. The highest accuracy ($66.41\%$), balanced accuracy ($66.44\%$), and $F_1$ score ($0.6704$) were obtained using the same representation with a linear SVM.

\begin{table}[htbp]
\centering
\small
\caption{
Subject-independent left-hand versus right-hand motor-imagery classification.
}
\label{tab:mi_results}
\setlength{\tabcolsep}{3pt}
\resizebox{\linewidth}{!}{%
\begin{tabular}{lllc cccc}
\toprule
\textbf{Optimized Metric}
& \textbf{Feature Config.}
& \textbf{Classifier}
& \textbf{Feat.}
& \textbf{AUROC}
& \textbf{Acc.}
& \textbf{B. Acc.}
& \textbf{$F_1$}
\\
\midrule

AUROC
& EEGSync-$\Delta$ + Statistical
& Ridge
& 16
& \textbf{0.7150}
& 66.17\%
& 66.18\%
& 0.6622
\\

\addlinespace

Acc. / B. Acc. / $F_1$
& EEGSync-$\Delta$ + Statistical
& Linear SVM
& 16
& 0.7147
& \textbf{66.41\%}
& \textbf{66.44\%}
& \textbf{0.6704}
\\

\bottomrule
\end{tabular}%
}
\end{table}

\begin{figure}[htbp]
    \centering

    \begin{subfigure}[t]{0.48\linewidth}
        \centering
        \includegraphics[width=\linewidth]{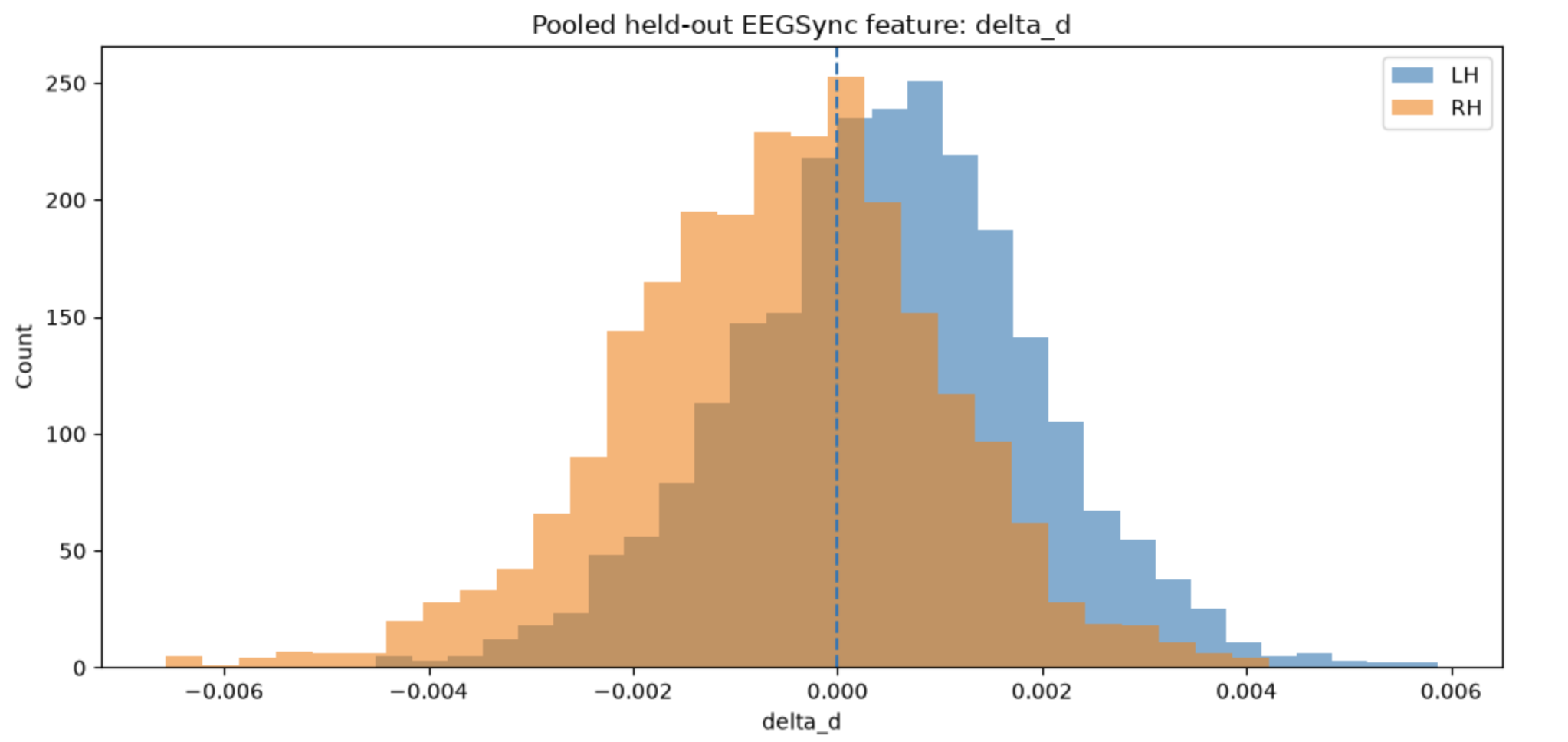}
        \caption{$\Delta d$ distribution.}
        \label{fig:mi_dist_deltad}
    \end{subfigure}
    \hfill
    \begin{subfigure}[t]{0.48\linewidth}
        \centering
        \includegraphics[width=\linewidth]{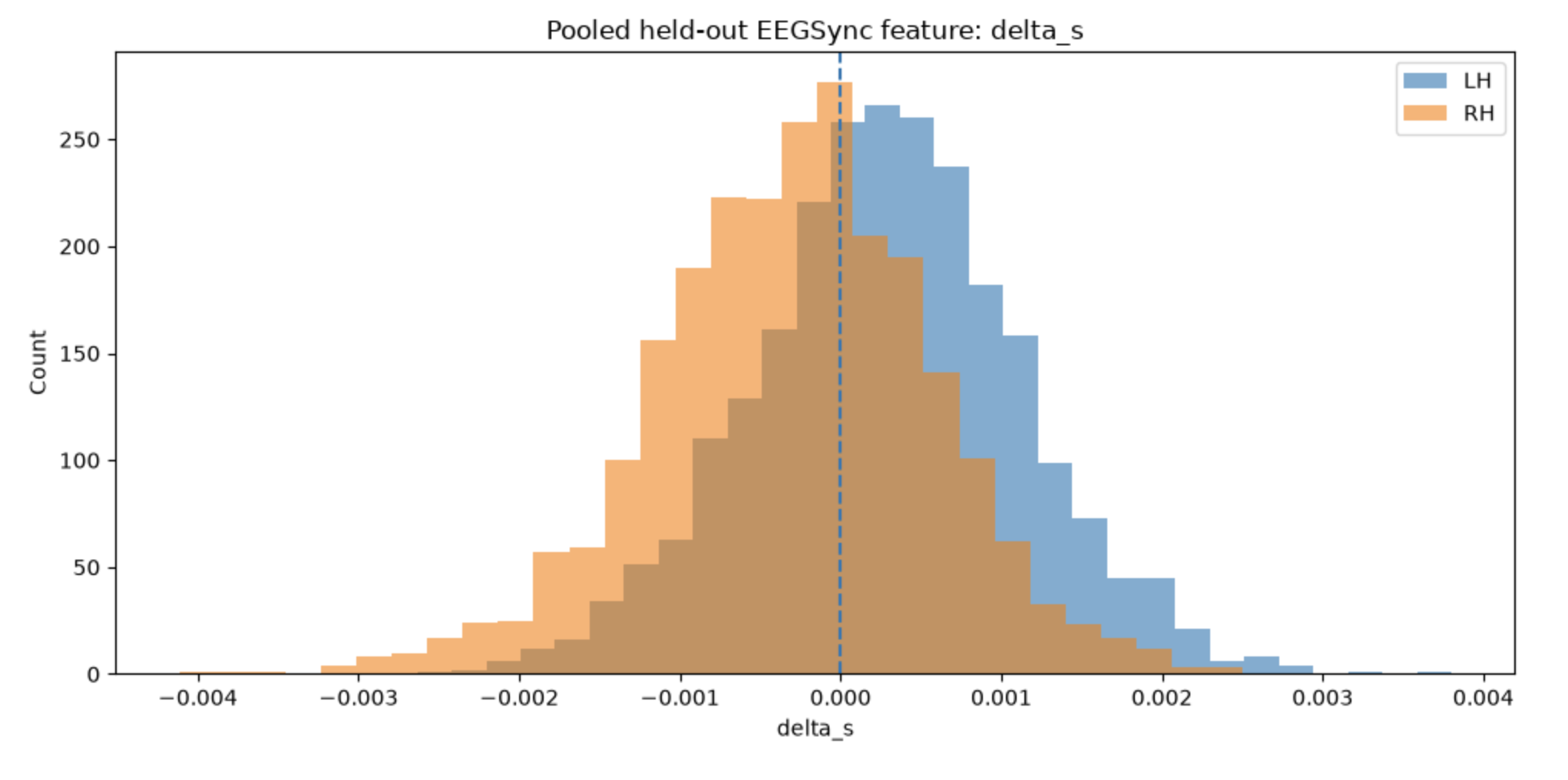}
        \caption{$\Delta s$ distribution.}
        \label{fig:mi_dist_deltas}
    \end{subfigure}

    \caption{
    \textbf{Held-out EEGSync feature distributions for left- and right-hand motor imagery.}
    Both synchronization measures exhibit task-dependent shifts, although with substantially greater overlap than in the primary resting-state validation experiments.
    }
    \label{fig:mi_dist}
\end{figure}

The held-out synchronization-feature distributions are shown in Fig.~\ref{fig:mi_dist}. Compared with the EO--EC and Alzheimer's disease experiments, the LH and RH distributions exhibit substantially greater overlap. Nevertheless, both $\Delta d$ and $\Delta s$ show systematic shifts between the two motor-imagery conditions.

Motor imagery therefore serves as a supplementary stress test rather than a primary validation task. The lower performance relative to EO--EC and Alzheimer's disease classification reflects the greater overlap between task classes, but the above-chance subject-independent performance indicates that EEGSync retains task-related information even in this more difficult setting.

\subsection{Numerical and Computational Validation }
\label{sec:numerical_validation}

We first verified that the economy-SVD implementation reproduces the direct EEGSync solution without approximation. Random complex matrices with $V=19$ channels and temporal dimensions ranging from $T=50$ to $T=800$ were aligned using both the explicit $T\times T$ unitary Procrustes solution and the compact formulation of Section~\ref{sec:efficient_eeg_sync}. Across all tested dimensions and trials, the resulting aligned representations agreed to numerical precision. The median relative Frobenius error was
\[
3.31\times10^{-15},
\]
with a maximum error of
\[
3.81\times10^{-15}.
\]
These results confirm that the economy-SVD implementation is a numerical reformulation of the same optimization problem rather than a low-rank approximation.

We next tested whether EEGSync preserves the pairwise inner products between EEG channels, as predicted by the analysis in Section~\ref{sec:geometry_preservation}. The channel inner-product matrix was preserved to numerical precision after alignment , with a median relative error of
\[
3.45\times10^{-15}
\]
and a maximum error of
\[
5.46\times10^{-15}.
\]
The singular spectrum was similarly unchanged, with a median relative error of
\[
9.13\times10^{-16}.
\]
This is expected because the squared singular values of $Z$ are the eigenvalues of $ZZ^\dagger$, which is preserved exactly under EEGSync alignment.

Finally, we compared the computational scaling of the direct and economy-SVD implementations for $V=19$ channels. At $T=100$, $250$, and $500$, the economy formulation was approximately $3.6\times$, $13.1\times$, and $36.6\times$ faster, respectively, than explicitly factorizing the temporal cross-product. The economy-SVD implementation continued to scale efficiently to longer recordings, requiring approximately $11.7$ ms at $T=10{,}000$ in the present benchmark. In contrast, the direct formulation requires construction and factorization of a $T\times T$ complex matrix, making its computational and memory requirements increasingly unfavorable as recording duration grows.

Together, these experiments verify the central implementation claims of EEGSync: the economy-SVD formulation reproduces the full unitary alignment to machine precision, preserves the predicted pairwise relationships between EEG channels, and substantially reduces the computational burden associated with long EEG recordings.

\begin{figure*}[H]
    \centering

    \begin{subfigure}[t]{0.32\textwidth}
        \centering
        \includegraphics[width=\linewidth]
        {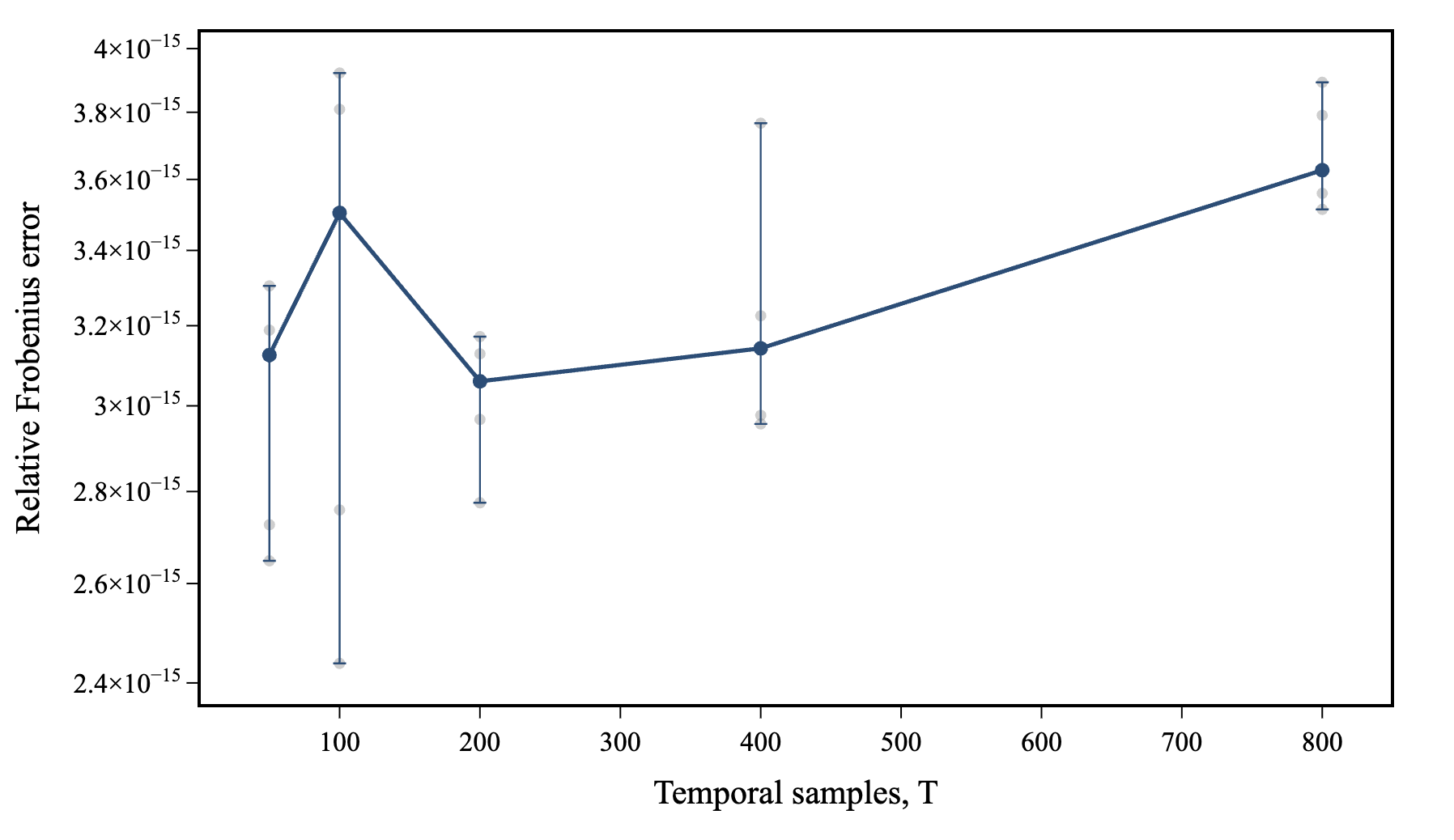}
        \caption{}
        \label{fig:validation_equivalence}
    \end{subfigure}
    \hfill
    \begin{subfigure}[t]{0.32\textwidth}
        \centering
        \includegraphics[width=\linewidth]
        {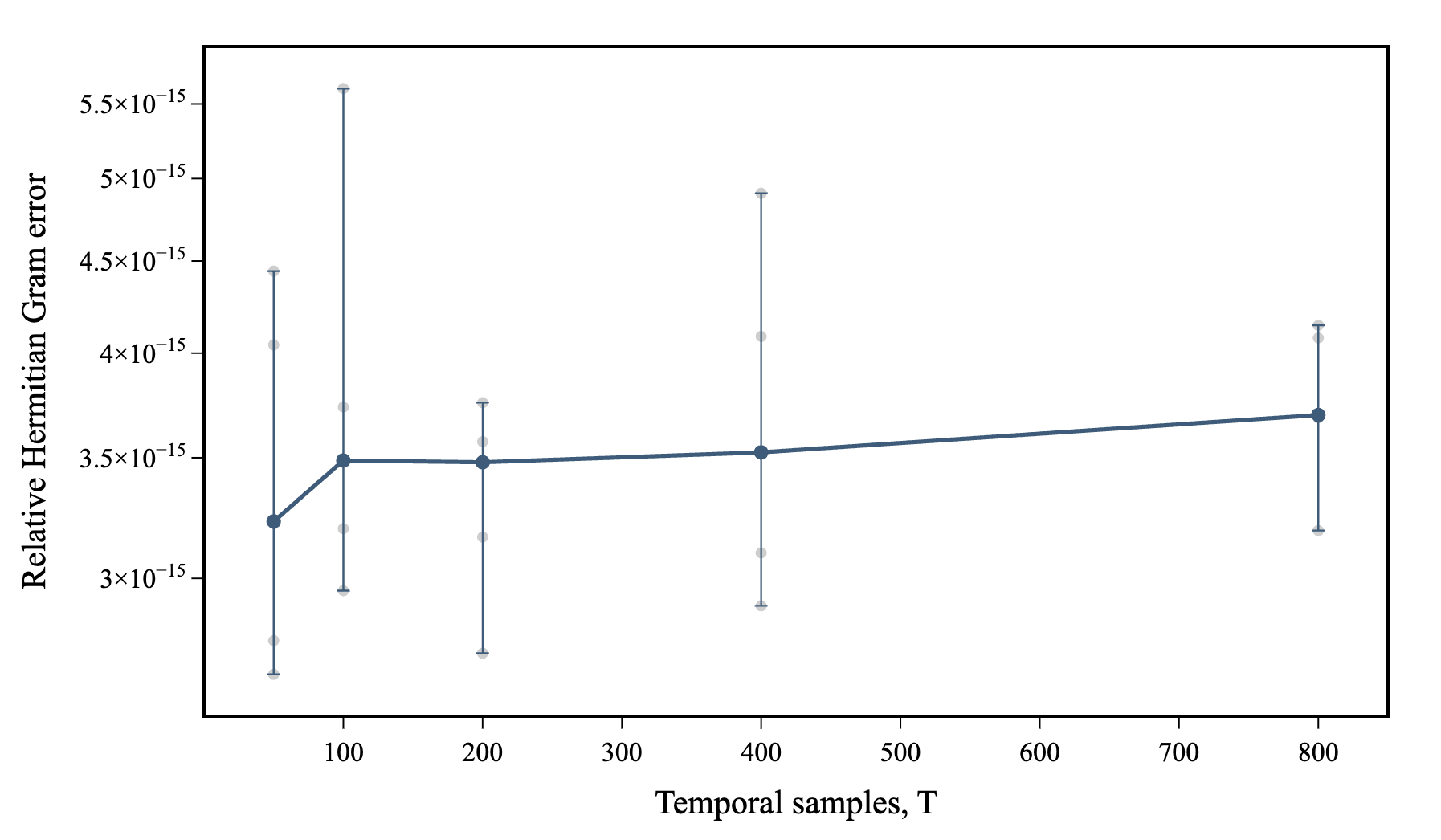}
        \caption{}
        \label{fig:validation_gram}
    \end{subfigure}
    \hfill
    \begin{subfigure}[t]{0.32\textwidth}
        \centering
        \includegraphics[width=\linewidth]
        {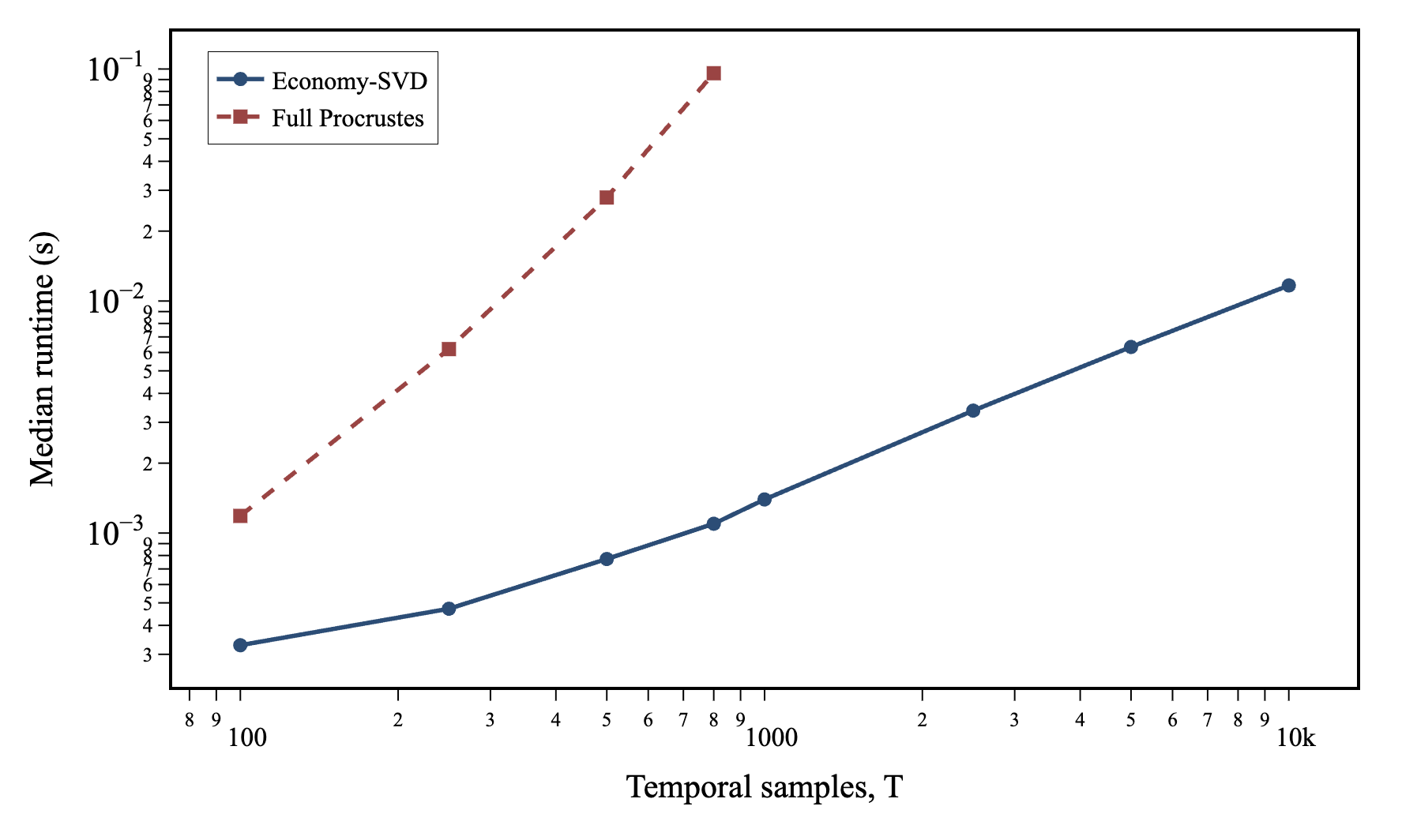}
        \caption{}
        \label{fig:validation_runtime}
    \end{subfigure}

    \caption{
    \textbf{Numerical and computational validation of EEGSync.}
    \textbf{(a)} Relative Frobenius error between the aligned
    representations produced by the direct unitary Procrustes solution
    and the economy-SVD implementation across temporal lengths $T$.
    Errors remain at machine precision, confirming that the compact
    formulation is an exact numerical reformulation rather than an
    approximation.
    \textbf{(b)} Relative error in the matrix of pairwise inner products between EEG channels
    before and after EEGSync alignment. Preservation to numerical
    precision confirms the invariant
    $(ZU)(ZU)^\dagger = ZZ^\dagger$.
    \textbf{(c)} Median runtime of the direct and economy-SVD
    implementations for $V=19$ channels. The computational advantage
    of the economy-SVD formulation increases rapidly with recording
    duration.
    }
    \label{fig:numerical_validation}
\end{figure*}

\end{document}